\documentclass[pre,twocolumn,superscriptaddress]{revtex4}

\usepackage{amsmath,amssymb}
\usepackage{graphicx}
\usepackage{bm}
\usepackage{placeins}
\usepackage{color}
\usepackage[colorlinks,linkcolor=blue,anchorcolor=blue,citecolor=blue]{hyperref}
\usepackage[T1]{fontenc}
\usepackage{verbatim}
\usepackage{multibib}

\begin{document}

\title{Machine-Learning Solutions for the Analysis of Single-Particle
Diffusion Trajectories}

\author{Henrik Seckler}
\affiliation{Institute for Physics \& Astronomy, University of Potsdam, 14476
Potsdam-Golm, Germany}
\author{Janusz Szwabi{\'n}ski}
\affiliation{Hugo Steinhaus Center, Faculty of Pure and Applied Mathematics,
Wroc{\l}aw University of Science and Technology, Wybrze{\.z}e Wyspia{\'n}skiego
27, 50-370 Wroc{\l}aw, Poland}
\author{Ralf Metzler}\email{rmetzler@uni-potsdam.de}
\affiliation{Institute for Physics \& Astronomy, University of Potsdam, 14476
Potsdam-Golm, Germany}
\affiliation{{Asia Pacific Center for Theoretical Physics, Pohang 37673,
Republic of Korea}}

\date{\today}

\begin{abstract}
Single-particle traces of the diffusive motion of molecules, cells, or animals
are by-now routinely measured, similar to stochastic records of stock prices
or weather data. Deciphering the stochastic mechanism behind the recorded
dynamics is vital in understanding the observed systems. Typically, the
task is to decipher the exact type of diffusion and/or to determine system
parameters. The tools used in this endeavor are currently revolutionized
by modern machine-learning techniques. In this Perspective we provide
an overview over recently introduced methods in machine-learning for
diffusive time series, most notably, those successfully competing in the
Anomalous-Diffusion-Challenge. As such methods are often criticized for their
lack of interpretability, we focus on means to include uncertainty estimates
and feature-based approaches, both improving interpretability and providing
concrete insight into the learning process of the machine. We expand the
discussion by examining predictions on different out-of-distribution data.
We also comment on expected future developments.
\end{abstract}

\maketitle

Single Particle Tracking (SPT) refers to the observation of the microscopic
motion of molecules. In 1828 Robert Brown used SPT to observe the movement of
granular particles, laying the foundations of Brownian Motion \cite{BRO28}.
After advancements in theory spearheaded by Einstein, Smoluchowski, Sutherland,
and Langevin, Jean Perrin was able to give a first estimate of Avogadro's
number by observing particle motion in a colloid \cite{PER09}. While SPT
applies mainly to observing the movement of molecules or micron-sized
tracer particles \cite{ELF19,CHE19,HOE13,HOR10,TOL04,LEI12,COD08,spt,spt1,
membr,membr1,membr2,membr3},
similar SPT data are also garnered in systems ranging from the movement
of animals \cite{OKU86,VIL22a,BAR05} to eye movement \cite{ENG11,eng} or stock
dynamics \cite{MAL99,PLE00}. Understanding such trajectories and developing
techniques for their analysis is thereby of vital importance in a multitude
of different fields \cite{FER13,AND76,OKU86,VIL22a,LUD11,COD08,MAL99,BOU03}.
Mathematically such a motion is described by a random walk as introduced by
Karl Pearson \cite{PEA05}. Here the position $x_i$ of a particle at time $t_i$
is obtained via a sequence of random steps $\Delta x_i$ $(i=1,...,T-1)$, such
that $x_n=x_0+\sum_{i=1}^{n}\Delta x_i$ $(n=0,...,T-1)$. In the simplest case
called "Wiener process", whose steps $\Delta x_i$ are independent and
identically distributed according to $(2\pi \sigma^2)^{-1/2}\exp\left(-\Delta
x_i^2/[2\sigma^2]\right)$ with constant waiting time $t_i-t_{i-1}=\Delta t$,
will lead to a Gaussian probability density function (PDF)
\begin{equation}
f(x,t)=\frac{1}{\sqrt{4\pi K_1t}}\exp\left(-\frac{x^2}{4K_1t}\right),
\end{equation}
where $K_1=\sigma^2/\Delta t$. Due to the action of the Central Limit
Theorem, the same PDF is reached as long as the increments are independent
and identically distributed with finite variance and finite mean waiting
time \cite{MIS19,MON65}. In particular this entails a linear growth of the
mean squared displacement (MSD) \cite{KAM92,LEV48,HUG95}
\begin{equation}
\langle x^2(t)\rangle\sim 2K_1t.
\end{equation}
This type of behavior is referred to as normal diffusion, the most well-known
example being the aforementioned Brownian Motion as described by Einstein,
Smoluchowski, Sutherland and Langevin when analyzing the motion of small
particles suspended in liquids or gases \cite{EIN05,SMO06,SUT05,LAN08}.

In practice however one often observes a non-Gaussian probability
density function and/or an MSD that grows non-linearly in time
\cite{BOU90,MET00,GOL06,MAN15,KRA19,STA17,KIN17,SOK12,HOE13,HOR10,TOL04,
LEI12,SAX94,SAX01,BUR11,ERN14}. Here we focus on the frequent case of
power-law growth of the MSD,
\begin{equation}
\label{eq_msd}
\langle x^2(t)\rangle\sim 2K_{\alpha}t^\alpha,
\end{equation}
referred to as "anomalous diffusion" with the anomalous diffusion
exponent $\alpha$. A growth slower than linear ($0<\alpha<1$) is called
subdiffusive, whereas a faster than linear growth ($\alpha>1$) is referred to
as superdiffusive, with the special case of ballistic motion for $\alpha=2$.
For such behavior to emerge, one or more of the conditions for the CLT to kick
in need to be violated, as is the case when the system shows heterogeneities,
long time correlations, diverging mean waiting times and/or infinite jump
variance. As an example, one may consider a granular gas with a temperature
changing over time, which causes non-identically distributed increments since
the increment variance is temperature dependent \cite{anna,anna1}. As a random
walk such a motion is modeled by Scaled Brownian Motion (SBM), in which the
diffusivity is time dependent \cite{LIM02,JEO14}. A diffusivity increasing
with time will lead to superdiffusion, while a decreasing one will lead
to subdiffusion. As another prominent example, long time correlations are
often observed in biomolecules, whose crowded environments lead to strong
anti-correlations (viscoelastic effects), while active motion may give
rise to strong correlations. Mathematically such motion is often modeled
as so-called Fractional Brownian Motion (FBM) \cite{MAN68a}. There exist
plenty of other models to explain the occurrence of anomalous diffusion
\cite{MET14,MAR22,HEG22,VIT22,SAB22,WAN20}, apart from the mentioned SBM and
FBM, we here also consider continuous-time random walk (CTRW) with random
waiting times in between successive jumps \cite{MON65,HUG81,WEI89}, L\'evy
walks (LWs) \cite{LEV37, CHE08, SHL86,ZAB15,MAG20} and annealed transient
time motion (ATTM) \cite{MAS14}. We provide short descriptions of each of
these models in the Supporting Information.

Since each of these models describes different physical causes for anomalous
diffusion, identifying the best fitting stochastic model is an important step
in unraveling the physical origin of an experimentally observed anomalous
diffusion \cite{MER15,CHE19,GOL06,MAN15,KRA19,STA17,KIN17}. Similarly
determining specific parameters attributed to each model, such as
the anomalous diffusion exponent $\alpha$ and coefficient $K_\alpha$,
can help quantify and/or differentiate between trajectories or systems
\cite{MAK11,GOL06}.  Typically this task is tackled through the use of
statistical observables, aiming at quantifying the expected differences between
the models \cite{MET09,MAG09,BUR11,MET19,VIL22b,SPO22,CON07,SLE19,MAR20,CHE17}.
However, the stochastic nature of these models in combination with the often
noisy and limited experimental data can severely hinder this process, and
may lead to conflicting results from different observables. For example,
it has been shown that noisy data can lead to a mistaken identification as
subdiffusion \cite{MAR02,SPO22}.

The rising computing power of modern processors have brought along a competing
approach. Machine learning (ML) has already shown wide applicability in
physical chemistry \cite{PRE20}, and is increasingly used in a variety
of fields from material science \cite{BAC19}, to medicine \cite{BAT20},
or quantum chemistry \cite{DRA20}. In particular, in recent years ML
has also been applied to anomalous diffusion dynamics seen in SPT data
\cite{GRA19,BO19,MUN20b,MUN20a,GAJ21}.  Here the task of finding the best
way to determine the underlying diffusion model and model parameters is left
to machines trained on simulated trajectories, either by directly feeding
into the machine the raw position data or by extracting relevant features
from the trajectories first.

After shortly discussing classical methods, we here focus on the
competing approaches utilizing ML, most notably those introduced during the
so-called "Anomalous Diffusion (AnDi) Challenge"~\cite{MUN20a,andijpa}. To
address the "Black Box problem" we present a deeper look into approaches
including uncertainty estimates as well as those relying on extracted
features. We present tests for the limits of both approaches when applied
to out-of-distribution data. We conclude with a discussion on benefits,
shortcomings and expected future developments of ML techniques to analyze
anomalous diffusion data.

\paragraph*{Classical approach.} The simplest path to the anomalous exponent
is given by direct calculation of the scaling exponent of the MSD, which,
given an ensemble of $N$ trajectories, is defined as
\begin{equation}
\langle x^2(t)\rangle=\frac{1}{N}\sum_{n=1}^N(x^{(n)}(t)-x^{(n)}(0))^2\sim2
K_{\alpha}t^\alpha.
\end{equation}
In experiments, one often relies on time-series analysis, utilizing the time
averaged MSD (TAMSD),
\begin{equation}    
\left<\overline{\delta^2(\Delta)}\right>=\frac{1}{\mathcal{T}-\Delta}\int_0^{
\mathcal{T}-\Delta}\left<\left[ x(t+\Delta)-x(t) \right]^2\right> dt,
\end{equation}
with observation time $\mathcal{T}$.  As long as the system is ergodic, the
TAMSD for sufficiently long $\mathcal{T}$ will convey the same information
as the MSD. However for anomalous diffusion this is often not the case, for
instance, when models feature diverging mean waiting times such as CTRW or
LW. This indicates that, when experimental conditions allow access to both
ensemble MSD and TAMSD, the possible difference between their behavior allows
one two differentiate between ergodic and non-ergodic models.

An alternative method is provided by the $p$-variation test
\cite{MAG09,BUR10,LOC20}. The sample $p$-variation is calculated using the
difference between every $m$th element of the trajectory,
\begin{equation}
V_m^{(p)}=\sum_{k=0}^{(T-1)/m-1}|x_{(k+1)m}-x_{km}|^p.
\end{equation}
Different models often show different behavior of the $p$-variation. For example, 
for FBM we have,
\begin{equation}
V_m^{(p)} \propto m^{p\alpha/2 -1},
\end{equation}
implying that, as a function of $m$, the $p$-variation increases for
$p>2/\alpha$ and decreases for $p<2/\alpha$. This is in contrast to, e.g.,
CTRW, where the $p$-variation will decrease for $p>2$ and increase for $p<2$,
regardless of anomalous exponent $\alpha$. Thus, calculating the $p$-variation
for different $p$ values can help differentiate between models or, for some
models, provide an estimate of the anomalous exponent $\alpha$. However,
static noise may compromise the $p$-variation output, as, e.g., tested
for subdiffusive CTRWs \cite{noise}.  Alternatively, it is also possible to
decompose the anomalous dynamics into the Moses $M$, Noah $N$, and Joseph $J$
scaling exponents (with $\alpha/2=J+L+M-1$), obtained from the scaling of
the cumulative absolute increments, the sum of the squared increments and the
rescaled range statistic. Each of these exponents corresponds to the violation
of one of the three conditions for the CLT \cite{MAN68b,CHE17,AGH21,MEY22}.

Another method is given through the use of the single-trajectory power
spectral density (PSD) \cite{MET19,VIL22b,SPO22},
\begin{equation}
S(f,\mathcal{T})=\frac{1}{\mathcal{T}}\left|\int_0^\mathcal{T}dte^{ift}x(t)
\right|^2.
\end{equation}
Of particular interest here is the so-called coefficient of variation,
\begin{equation}
\gamma(f,\mathcal{T})=\frac{\sigma(f,\mathcal{T})}{\mu(f,\mathcal{T})},
\end{equation}
where $\sigma(f,\mathcal{T})$ and $\mu(f,\mathcal{T})$ are the mean value
and variance of the PSD. In FBM for example, $\gamma(f,\mathcal{T})$
shows distinct behavior for subdiffusion ($\gamma(f,\mathcal{T})\sim1$),
superdiffusion ($\gamma(f,\mathcal{T})\sim\sqrt{2}$) and normal diffusion
($\gamma(f,\mathcal{T}) \sim\sqrt{5}/2$), in the limit of high frequencies
or long observation times \cite{KRA19}. Single-trajectory PSDs are also
quite robust against static and dynamics noise \cite{SPO22}.

The aforementioned methods cover only a fraction of possibilities, other
techniques not further specified here include the use of the velocity
autocorrelation \cite{BUR11}, the first passage statistics \cite{CON07}, the
codifference \cite{SLE19} or the autocovariance \cite{MAR20}.  The applications
of these statistical techniques, however, struggle when data is sparse and
often require an ensemble of trajectories. For some of these methods, noisy
trajectories may also present a problem, as they may lead to a erroneous
identification, e.g., as subdiffusion \cite{MAR02,SPO22}.

As an alternative approach to classify SPT data Thapa et al.\ demonstrated,
that Bayesian Inference may be used to determine the best fitting model
and its parameters directly from the position data of an SPT experiment
using their mathematical description \cite{THA18}. Specifically, these
descriptions allow one to directly calculate the likelihood of a given
trajectory for a specific model with given parameters. These parameters are
then adjusted to maximize the probability of the trajectory. The difference
in maximum likelihood is used to determine the most probable model. This
method has shown great promise for processes, for which the likelihood
is easily calculated in closed form, such as FBM or SBM. They struggle,
however, when models feature hidden waiting times, though there have been
recent advances using hidden Markov processes \cite{PAR21}. Even so, high
computational cost remains an issue of Bayesian Inference, often resulting
in a trade-off between computational feasibility and accuracy.

\paragraph*{The Anomalous Diffusion Challenge.} ML in recent years has grown
into a strongly competing class of approaches. In 2019 Granik et al., using
a convolutional neural network, demonstrated that one can differentiate
between simulated Brownian motion, CTRW and sub- or superdiffusive FBM
trajectories \cite{GRA19}. In the same year Bo et al.\ used a similar
procedure to determine the anomalous diffusion exponent of FBM trajectories
via a recurrent neural network \cite{BO19}. Similarly, in 2020 Mu{\~n}oz-Gil
et al.\ demonstrated that a random tree forest can differentiate between CTRW,
LW, FBM and ATTM, and provide an estimate for the anomalous diffusion exponent
\cite{MUN20b}. In all these cases, it was shown that ML can achieve better
accuracy than conventional methods, especially when the available data is
sparse. It should be noted, however, that these approaches all suffer from
the often-quoted "Black Box problem", outputting answers without explanations
as to how these are obtained \cite{SZE14}, as detailed below.

Among ML approaches, the mentioned strategies utilizing convolutional neural
networks \cite{GRA19,KOW19}, recurrent neural networks \cite{BO19,GAJ21},
and random tree forests \cite{MUN20b} already differ significantly. In
an effort to compare the performance of different techniques, in 2020
Mu{\~n}oz-Gil et al.\ launched the AnDi-Challenge \cite{MUN20a,andijpa}.
Reported in 2021, the goal of the AnDi-Challenge was to provide a competitive
comparison of different available methods to decode anomalous diffusion
\cite{MUN20a,MUN21a}. The AnDi-Challenge also continues to serve as a
benchmark to quickly assess the performance of newly developed or improved
methods \cite{ARG21,GAR21,LI21,FIR23,ALH22,GEN21,MAN21,KOW22,PIN21,
SEC22,AGH21,MEY22,PAR21,THA22}.

The challenge consisted of three tasks: (i) inference of the anomalous
diffusion exponent, (ii) classification of the diffusion models and (iii)
segmentation of trajectories. For tasks (i) and (ii), participants were given
a set of trajectories, each randomly generated from one of five different
anomalous diffusion models with a randomly chosen anomalous diffusion
exponent.  For task (iii), the model and/or exponent changed at a given
point in the trajectory. Participants were required to predict the change
point in addition to the model and the anomalous diffusion exponent in both
segments. To emulate experimental data, all trajectories were corrupted by
white Gaussian noise of varying strength. Moderately-sized training data
sets as well as the code necessary to generate further labeled data are
freely available in a repository \cite{MUN21b}.

In total, 15 teams participated in the AnDi-Challenge, using a variety
of different methods. While most teams used some form of ML, the more
traditional approaches were represented by teams using Bayesian Inference
\cite{KRO18,PAR21,THA22} and scaling analysis as well as feature engineering,
primarily based on a decomposition method using the Moses, Noah, and Joseph
exponents \cite{AGH21,MEY22}.

Several different ML techniques were used, some of which were applied
to the raw position data \cite{ARG21,BO19,GAR21,LI21,FIR23,ALH22}.
Others methods relied on features extracted from the input trajectories
\cite{GEN21,MAN21,JAN20,KOW19,KOW22,LOC20,PIN21}, or used a combination of
both strategies \cite{VER21,VER22}. The techniques using raw data focused on
deep learning (DL) \cite{GRA19,LI21,GAR21,FIR23,ALH22,ARG21,BO19}, while the
feature-based methods also included other ML methods such as gradient boosting
\cite{JAN20,KOW19,KOW22}, random forests \cite{JAN20,KOW19, KOW22,LOC20},
and extreme learning machines \cite{MAN21}. In the AnDi-Challenge the ML
methods outperformed the classical approaches with top results obtained by DL,
achieving an accuracy of 88\% for model classification and a mean absolute
error (MAE) of 0.14 for the regression of the anomalous diffusion exponent for
2D trajectories \cite{MUN21a}. For comparison, the more traditional Bayesian
Inference---with the limited amount of processes for which the likelihood
function was derived at that point---achieved an accuracy of 53\% and MAE
of 0.20 in the challenge \cite{MUN21a}. Classical observables such as the
above-mentioned decomposition method using the scaling exponents $M$, $N$,
and $J$, scored 51\% accuracy with an MAE of 0.31 \cite{MUN21a}.

\paragraph*{The raw data approach of deep learning.}

\begin{figure}
\includegraphics[width=\linewidth]{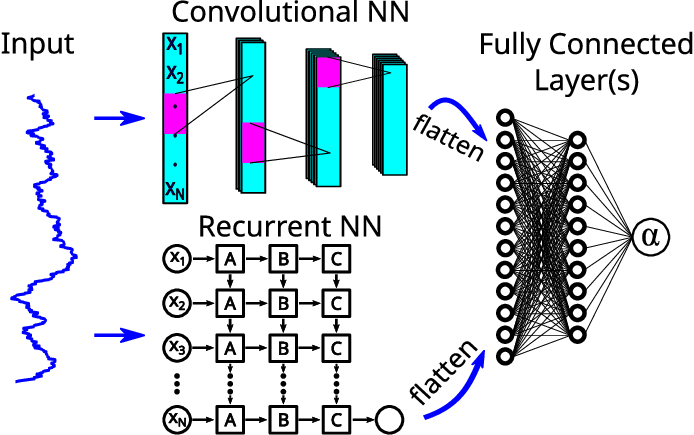}
\caption{Schematic representation of convolutional and recurrent neural
network architectures for the analysis of single particle trajectories.
In both cases input data consist of normalized trajectory positions
(or increments) $x_1,...,x_N$. In a convolutional neural network a kernel is
slid along the input data, generating outputs for each region. Usually each
layer consists of multiple kernels with identical sizes but different weights,
each generating a new data sequence depicted in the figure as an additional
dimension. We here show a convolutional neural network with three layers
utilizing 3, 5 and 6 kernels, respectively. In a recurrent neural network,
as depicted in the lower half, the data is passed in sequence through a
recurrent unit, with the output of the previous time step included as input
in the next step (vertical connections in the figure). Here we depict a stacked
recurrent neural network consisting of three layers with weight matrices
\textit{A,B,C} respectively. For both recurrent and convolutional networks
the resulting output is usually flattened into a one-dimensional array and
passed through one or multiple fully connected layers, ending, e.g., in a
prediction of the anomalous diffusion exponent $\alpha$ or diffusion
coefficient $K_{\alpha}$.}
\label{fig_deeplearn}
\end{figure}

Following the rising availability of high computational power along with
increasingly more detailed data sets, more and more ML approaches rely
on highly complex architectures involving thousands of parameters. With
DL we refer to neural networks with many hidden layers, often resulting in
several hundreds of thousand of fitting parameters (weights) \cite{LEC15}. The
complexity of these models allows them to directly learn from massive amounts
of raw data, with little to no need for human-engineered preprocessing.
Specifically for the analysis of anomalous diffusion, this entails directly
learning from the position time series of the recorded trajectories. To
speed up training and reduce the required data volume, the input data
will undergo minimal preprocessing, via a normalization of its standard
deviation. Since diffusion models only rely on the increments of a
process and no additional relevant information is included in the absolute
positions, the trajectories are often also converted to the increment process
\cite{MUN21b,ARG21,BO19,GAR21,LI21,FIR23,ALH22}. In inhomogeneous, static
environments, this condition may, of course, no longer hold.

The DL solutions, as presented in the AnDi-Challenge and newly developed
ones since, mostly utilize convolutional \cite{GRA19,LI21,GAR21,FIR23,ALH22}
and recurrent \cite{ARG21,BO19,GAR21,CHE22} neural network architectures. In
convolutional neural networks, best known for their applications in image
classification, the layers consist of one or several convolutional kernels
that are slid along the input tensor \cite{FUK80}. By stacking multiple
such layers, they are able to detect correlations in the sequence. Recurrent
neural networks, most notably the so called "long short-term memory" (LSTM)
networks, are specifically designed for time sequence data, making them useful
for tasks such as speech recognition, translation, or sequence forecasts
\cite{HOC97}. Layers typically consist of a single recurrent unit applied
successively to each time step, with outputs of the previous time step included
as additional inputs for the next time step. Figure \ref{fig_deeplearn} shows
simplified schematic representations of both architectures. Other notable
architectures that have been shown to be applicable to single particle
tracking data analysis include graph neural networks \cite{VER21,VER22}
and transformer/encoder networks \cite{FIR23,LI21}.

In the AnDi-Challenge all top results were achieved by DL
\cite{MUN21a,ARG21,GAR21,GRA19,LI21,GEN21}, though notably one of these did
rely on extracted features rather than the raw trajectories \cite{GEN21}.
Overall, DL, and ML in general, showed great promise in the community
challenge. However, one should not dismiss the shortcomings of such
methods, which are most often criticized for their lack of interpretability
\cite{SZE14}. To that end, we discuss added uncertainty estimations as well
as feature-based approaches in the following.

\paragraph*{Qualifying deep learning by including uncertainties.}
\label{sec_bdl}

\begin{figure}
\centering
\includegraphics[width=\linewidth]{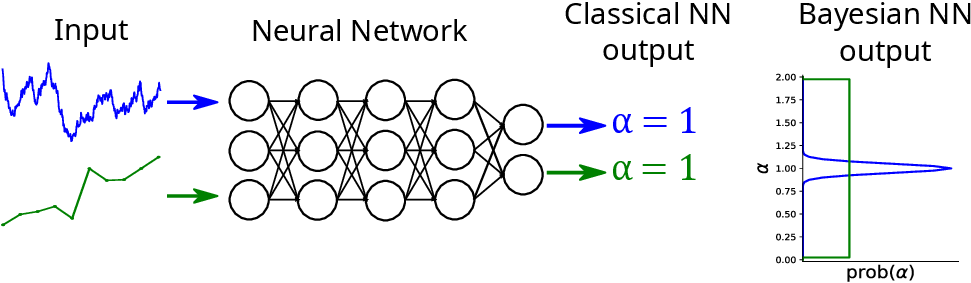}
\caption{Uncertainty problem of DL. The same anomalous diffusion exponent is
predicted for two trajectories of different length, when fed into a classical
neural network---despite their differing amount of information. The two cases
can only be distinguished when the probability distribution of possible output
anomalous diffusion exponents is considered, instead of a point estimate. Such
an estimate can be provided by Bayesian neural networks. Figure adapted
from \cite{SEC22}.}
\label{fig_uncert_prob}
\end{figure}

\begin{figure*}
\centering
\includegraphics[width=0.7\linewidth]{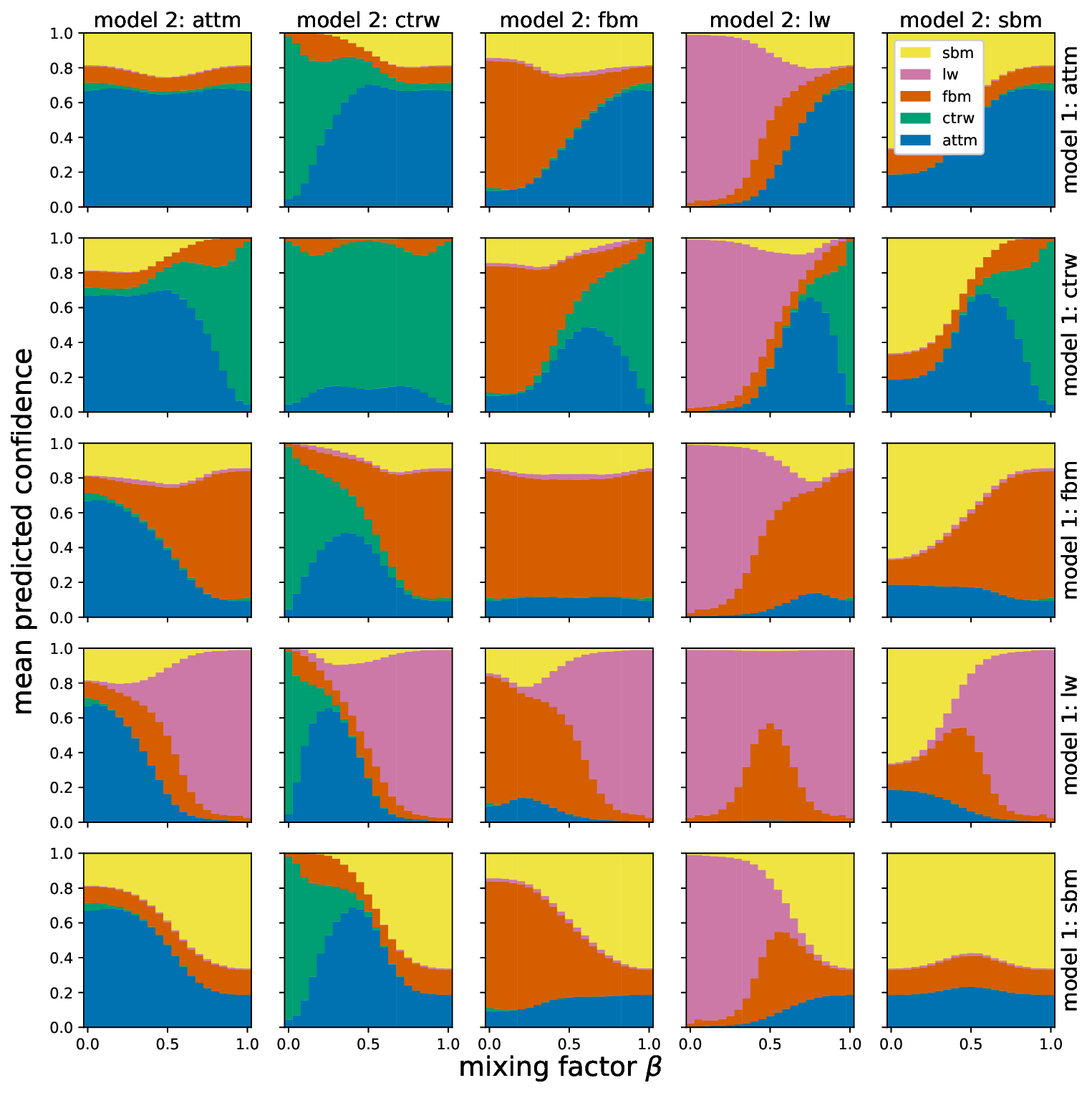}
\caption{ML-classification for a superposition of 2 models. The panels
depict the mean confidence assigned by the neural network, when presented
with a mixture of two models in dependence of the mixing factor $\beta$. The
depicted results are obtained from two-dimensional trajectories with 100
data points each.}
\label{fig_superpos_2d}
\end{figure*}

\begin{figure}
\centering
\includegraphics[width=0.55\linewidth]{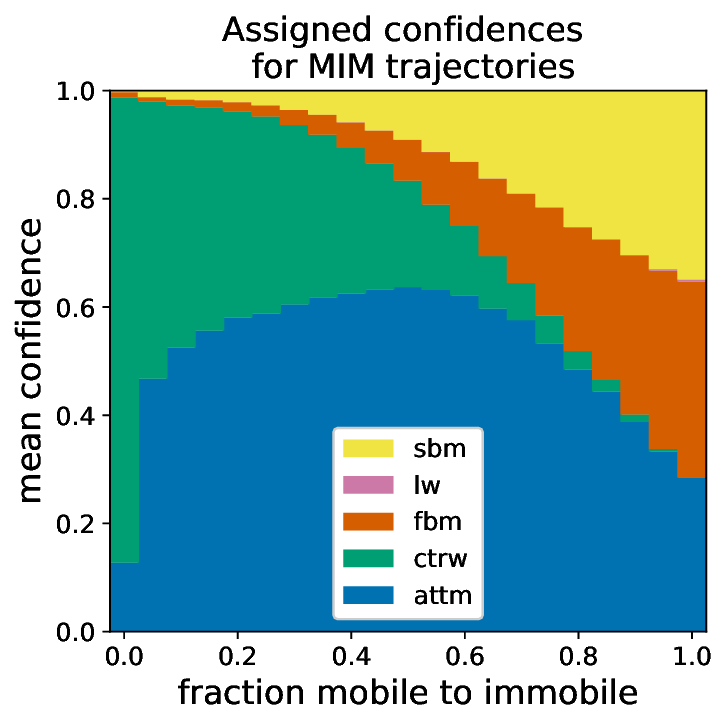}
\caption{Classification for Mobile-Immobile-Model trajectories for different
fractions of mobility. The MIM effectively converges to CTRW for low mobile
fractions $f_m$, and to Brownian motion for $f_m\approx1$. The depicted results
are obtained from one-dimensional trajectories with 250 data points each.}
\label{fig_mim_1}
\end{figure}

Classical DL models only provide point estimates of the output and
do not furnish any concrete information on the reliability of this estimate.
In extreme cases, this also means that these methods will provide outputs
on data, that have nothing to do with the learned problem. Even within
the desired task, different inputs may provide the same point estimate,
but underlie massively different uncertainties. As an example, figure
\ref{fig_uncert_prob} depicts two diffusion trajectories, which, fed into a
neural network, would both be assigned an anomalous diffusion exponent of
$\alpha=1$, even though one input contains considerably more information
(data points) than the other. To reveal the difference, one would need to
output a probability distribution of $\alpha$ values instead. Examining
such a distribution reveals that, while the first trajectory is roughly a
Brownian motion, the prediction of the second is just the result of obtaining
no relevant information, with the uniform distribution $\alpha\in[0,2]$
leading to a point estimate of $\alpha=1$.

To change the predictions to a probability distribution, we need to model
two types of uncertainty \cite{KIU09,KEN17}. \textit{Aleatoric} uncertainty
refers to the uncertainty inherent in the data, caused, for instance, by
measurement noise or an inherent stochasticity of the system. This uncertainty
remains even for a perfect model obtained from an infinite amount of data
and therefore must be included in the output of the neural network model and
trained by utilizing an appropriate loss function \cite{WAN22,NIX94}. As
no model is perfect, it is insufficient to consider aleatoric uncertainty
alone. Namely, to account for the difference between training and test
data, or an insufficient amount of training data in the first place,
one needs to introduce a second uncertainty measure. \textit{Epistemic}
(or \textit{systematic}) uncertainty can be included by considering the
weights of the neural network themselves as uncertain quantities. Formally,
the probability $p(\theta|\mathcal{D})$ of the weights $\theta$, given data
$\mathcal{D}$, is given by Bayes' rule \cite{KOL50},
\begin{equation}
\label{eq_bayes}
p(\theta|\mathcal{D})=\frac{p(\mathcal{D}|\theta)p(\theta)}{p(\mathcal{D})}.
\end{equation}
To obtain the final probability $p(y|x_i,\mathcal{D})$ for some output $y$
given the input $x_i$, we combine the aleatoric uncertainty, represented
by the probability $p(y|x_i,\theta)$ for one set of weights $\theta$,
with the epistemic uncertainty by marginalization over the weights.
The resulting integral is usually approximated through Monte Carlo sampling
\cite{MET49,binder},
\begin{eqnarray}
\nonumber
p({y}|x_i,\mathcal{D})&=&\int d\theta p({y}|x_i,\theta) p(\theta|\mathcal{D})\\
&\approx&\frac{1}{M}\sum_{m=1}^{M}p({y}|x_i,\theta_m),
\end{eqnarray}
where $\theta_m$ is sampled from the distribution $p(\theta|\mathcal{D})$
for a sufficient number $M$ of discrete points. As an exact calculation
of $p(\theta|\mathcal{D})$ (via equation \ref{eq_bayes}) quickly becomes
computationally infeasible for deep neural networks, one uses approximations
to generate the samples $\theta_m$. Various methods, summarized under the term
\textit{Bayesian Deep Learning}, have been proposed, the simplest of which
is to train an ensemble of neural networks, known as \textit{deep ensembles}
\cite{LAK17}. Other ways to generate samples include \textit{MC-Dropout}
\cite{GAL16a,GAL16b}, in which one uses dropout to generate multiple samples
from the same neural network, and \textit{Stochastic Weight Averaging Gaussian}
(SWAG) \cite{MAD19,WIL20}, which approximates $p(\theta|\mathcal{D})$ by
a Gaussian distribution, obtained by interpreting a stochastic gradient
descent \cite{BOT10} as an approximate \textit{Bayesian Inference} scheme.

Recently it was demonstrated that based on \textit{Multi-SWAG}, a combination
of SWAG and deep ensembles, one can add informative uncertainty predictions
to the DL solution for the analysis of single-particle anomalous diffusion
trajectories \cite{SEC22}. The introduced method maintains the performance
of the top AnDi-Challenge competitors, while it provides a well calibrated
uncertainty estimate with expected calibration errors \cite{NAE15,LEV20} of
only 0.0034 for the regression of $\alpha$ and 0.45\% for the classification
of the diffusion model. On top of this, it was demonstrated \cite{SEC22}
that the added error prediction improves the interpretability of the deep
neural networks, demonstrating in detail that the predicted behavior can be
linked to properties of the underlying diffusion models. In the Supplementary
figure S1 an example is shown of how error predictions can be analyzed when
inferring the anomalous diffusion exponent.

To further elaborate on the study in \cite{SEC22}, we now discuss the results
obtained from the Multi-SWAG approach when confronting the introduced networks
with previously unseen out-of-distribution data. First, we examine the outputs
when feeding the network with a superposition of two models, the increments
of which are obtained by the weighted sum of the increments of two models
with random anomalous diffusion exponents. With the mixing factor $\beta$
we then obtain
\begin{equation}
\Delta x_{\text{new}}=\beta\Delta x_{\text{model1}}+(1-\beta)\Delta x_{\text{
model2}}.
\end{equation}
Based on two-dimensional trajectories of length 100, figure
\ref{fig_superpos_2d} shows the mean confidences, that is, the mean value of
the predicted model probabilities over $2\times10^5$ input trajectories, in
dependence of the mixing factor for different model combinations, represented
by the rows and columns in the panel grid. For the convenience of the reader,
the panels include the redundant case of swapped model 1 and 2, which results
in a symmetry with respect to the panel grid diagonal, i.e., superpositions
of a model with itself. In most cases we see a smooth transition of the
confidence from the marginal cases on the left and right, which are the
normal predictions for pure trajectories of model 2 and 1, respectively. A
notable exception, however, is the behavior for superpositions with CTRW, as
these often show high probabilities for ATTM. Since ATTM could be considered
a combination of CTRW and Brownian motion, often showing the jumping motion
of CTRW interspersed with Brownian motion, this is not unexpected. Moreover,
we see that superpositions with LW often show high probabilities for FBM,
which can be explained due to the similarity of LW with highly correlated
FBM. Analogous behavior can be seen in 1D in the Supplementary figure S2.

As another example we confront the trained neural network with trajectories
obtained from the "Mobile-Immobile-Model" (MIM) \cite{DOE22,GEN76, DEA63}. In
the MIM, trajectories switch between a mobile and an immobile state, with
mean residence times $\tau_{\text{m}},\tau_\text{im}$. At equilibrium the
fraction of time a test particle spends in the mobile phase is given by
$f_m=\tau_m/(\tau_m +\tau_{im})$. This model provides information about
the immobilized fraction of the particle motion. Moreover, it includes
a continuous transition between a normal-diffusive ($\alpha=1$) CTRW on
the one side, for a low fraction of the mobility ($f_m\rightarrow0$), and
Brownian motion on the other side, for a high fraction of the mobility
($f_m\rightarrow1$). The results, depicted in figure \ref{fig_mim_1},
confirm that the method correctly classifies the two extremes as CTRW for
low mobility and as Brownian motion for high mobility, which for this method
is represented as a split probability between SBM, FBM and ATTM (all three
models that can exhibit Brownian motion). In between these two limits we
see high confidences for ATTM, which is not surprising, as ATTM is the only
model, of those considered here, that mimics the phase switching behavior
of a MIM trajectory.

\paragraph*{Feature-based classification of single particle trajectories.}
As demonstrated by the AnDi-Challenge \cite{MUN21a}, DL methods perform
excellently in the analysis of the diffusion models and outperform the
more traditional approaches to single particle tracking data. However, the
choice of a suitable classification method is usually more subtle than simply
looking at its performance. The availability of tools and libraries for DL
makes it relatively easy to quickly create effective predictive models. But
due to their complexity, those models are Black Boxes providing (almost)
no insight into the decision making processes. In the previous section
we showed how confidences can be established to judge the validity of the
provided output. Here we consider the interpretability of the parameters
in the ML approach. To give an example for the complexity in DL consider
ResNet18, one of the simplest deep residual network architectures used in
\cite{GAJ21} for trajectory classification. This network originally had
11,220,420 parameters. The authors were able to reduce this number down
to 399,556 with a positive impact on accuracy of the resulting classifier.
Although this is an impressive achievement, the interpretation of all those
remaining parameters is, of course, practically elusive.

\begin{figure*}
\centering
\includegraphics[scale=0.55]{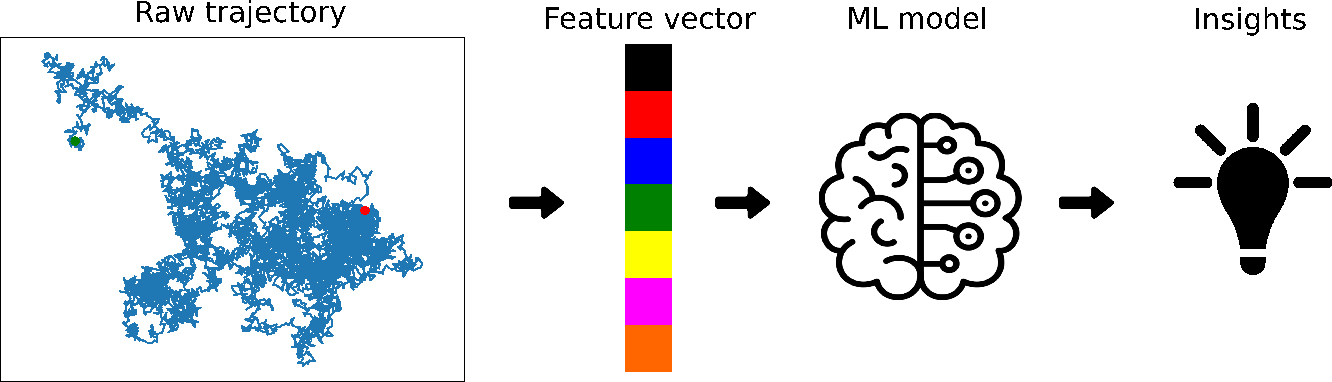}
\caption{Schematic workflow of the feature-based method: a set of features is 
extracted from raw trajectories and used as input to the classification or  
regression model. Analysis of the impact of the features on the outcome give
insights into the decisionmaking process of the model.}
\label{fig_feature-based}
\end{figure*}

The tradeoff between model accuracy and its interpretability is one of the
reasons for feature-based attempts for the classification of diffusion models
\cite{WAG17,KOW19,MUN20b,JAN20,LOC20,KOW22}. These feature-based methods are
statistical learning algorithms that do not operate on raw data. Instead,
each data sample is characterized by a vector of human-engineered features or
attributes. Those vectors are then used as input for a classifier (see figure
\ref{fig_feature-based} for a workflow of the method). In some sense, those
methods may be treated as a kind of extension to the statistical techniques
usually used for classification purposes. Instead of conducting a testing
procedure based on one statistics, we can turn all of them into features
and use them to train the model. This could be of particular importance in
situations, when single statistics yield inconclusive results or when testing
results based on different statistics significantly differ from each other
[87]. Automated feature-based analysis can thus be used in addition to deep
learning methods to learn more about the values of specific features and their
relative importance in categorizing input data.

Feature engineering, i.e., the process of extracting attributes from raw
data, is not a trivial task. It requires domain expertise to pinpoint which
features may be valuable for the process that generated the given set of
data. It may also be time and resource consuming, as testing the impact of
newly created features on the predictions involves repetitive trial and error
work. It has been already shown \cite{KOW19} that classifiers, which were
trained with a popular set of features, may not generalize well beyond the
situations encountered in the training set. Thus, careful attention needs
to be paid to the choice of the attributes. They should cover all important
characteristics of the process, but, at the same time, they should contain
the minimal amount of unnecessary information, as each redundant piece of
data causes noise in the classification and may lead to overfitting (see
\cite{JAM13} for a general discussion concerning the choice of features).

\begin{figure}
\centering
\includegraphics[scale=1]{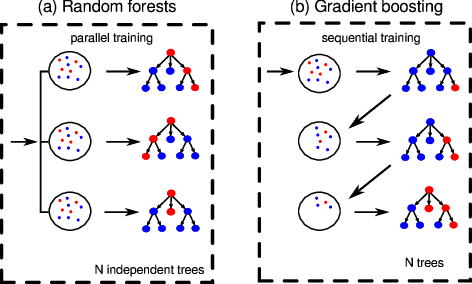}
\caption{Schematic comparison between random forest (left) and gradient
boosting methods (right). In the random forest, $N$ independent trees are
built in parallel from random subsets of the input data set. In gradient
boosting, the next tree is constructed from the residuals of the ensemble
and added to it.}
\label{fig_ensemble}
\end{figure}

Once the appropriate set of features is identified, the choice of an
actual classification algorithm is of secondary importance. Very often,
random forest \cite{WAG17,KOW19,MUN20b,JAN20,LOC20} or gradient boosting
\cite{KOW19,MUN20b,JAN20,LOC20,KOW22} methods are used, because they
offer a reasonable compromise between the accuracy of the results and their
interpretability. Both algorithms fall into the category of ensemble learning,
i.e., methods that generate many classifiers and aggregate their results. In
both cases, decision trees \cite{SON15} are used as the basic classifier. In a
random forest, several trees are constructed from the same training data. For a
given input, the predictions of individual trees are collected and then their
mode is taken as the output. In case of gradient boosting, the trees are not
independent. Instead, the single classifiers are built sequentially from the
mistakes committed by the ensemble (see figure \ref{fig_ensemble}). In terms
of interpretability, both algorithms are placed between single decision trees
(that are easy to interpret) and DL (with the Black Box problem).

In the AnDi-Challenge, the feature-based contribution was outperformed by the
winning teams using DL (73\% accuracy versus 88\% for the winners).  However,
the authors of the feature-based method further elaborated on their set of
features to achieve 83\% accuracy on the same validation set \cite{KOW22}.
This was based on a mixture of characteristics tailor-made to the diffusion
processes (e.g., MSD, anomalous diffusion exponent, diffusion coefficient)
and problem-agnostic ones (e.g., detrending moving average, kurtosis). All
features used in \cite{MUN20a} and \cite{KOW22} are summarized in table
\ref{tab_features}.

\begin{table}
\centering
\begin{tabular}{p{6cm}}
\hline\hline
\textbf{Original features}\\\hline\hline
Anomalous exponent\\\hline
Diffusion coefficient\\\hline
Asymmetry\\\hline
Efficiency\\\hline
Empirical velocity autocorrelation function\\\hline
Fractal dimension\\\hline
Maximal excursion\\\hline
Mean maximal excursion\\\hline
Mean Gaussianity\\\hline
Mean-squared displacement ratio\\\hline
Kurtosis \\
\hline
Statistics based on $p$-variation\\\hline
Straightness\\\hline
Trappedness\\\hline\hline
\textbf{Additional features}\\\hline\hline
D'Agostino-Pearson test statistic\\\hline
Kolmogorov-Smirnov statistic against $\chi^2$ distribution\\\hline
Noah exponent\\\hline
Moses exponent\\\hline
Joseph exponent\\\hline
Detrending moving average\\\hline
Average moving window characteristics\\\hline
Maximum standard deviation\\\hline\hline
\end{tabular}
\caption{Features used to characterize single particle trajectories. The
original set of features was used in the AnDi-Challenge and achieved 73\%
accuracy. With the additional features, the performance of the classifier
increased to 83\%. The definitions of the features may be found in
Appendix B and in \cite{KOW22}.}
\label{tab_features}
\end{table}

The authors were able to assess the importance of the features in the overall
classification and to calculate the contribution of each attribute to the
classification of every single trajectory, giving some insight into the
decision making process of the classifier \cite{KOW22}. The results achieved
with a simple gradient boosting method indicate that the feature-based ML,
overshadowed somewhat by DL approaches in the recent years, constitutes a
serious alternative to the state-of-the-art approaches.  It should be also
mentioned that better interpretability is not the only benefit related to
feature-based methods. Compared to DL, they usually work better on small
data sets and are computationally (and thus also financially) cheaper,
see \cite{KOW19} for a short comparison. Additionally, in case of single
particle tracking data, they naturally allow for the simultaneous analysis
of trajectories of different lengths.

\paragraph*{Testing the limitations of machine learning.}

\begin{table}
\begin{center}
\begin{tabular}{c|c|c|c|c}
dynamic noise $n_e$ & \multicolumn{2}{c|}{MAE} & \multicolumn{2}{c}{accuracy} \\
& DeepL & feature & DeepL & feature\\
\hline
1 & 0.207 & 0.23 & 78\% & 71\% \\
2 & 0.221 & 0.23 & 69.8\% & 71\% \\
5 & 0.229 & 0.22 & 59.3\% & 68\% \\
10 & 0.232 & 0.22 & 55.2\% & 65\% \\
20 & 0.235 & 0.22 & 53.5\% & 65\%\\
\hline
\end{tabular}
\end{center}
\caption{Performance of ML models, when confronted with data corrupted
by dynamic noise of different strength, as characterized by the exposure
length $n_e$. The case $n_e=1$ corresponds to no dynamic noise. The table
shows the performance for the DL-based method introduced in \cite{SEC22}
as well as a feature-based method utilizing the features introduced in
\cite{KOW22}.}
\label{tab_dynnoise}
\end{table}

\begin{table}
\begin{center}
\begin{tabular}{l||c|c}
\hline\hline 
& \multicolumn{2}{c}{MAE} \\ 
\cline{2-3} 
& NN50 & NN200 \\ 
\hline \hline
deep learning model & 0.246 & 0.264 \\ 
\hline 
feature-based model & 0.348 & 0.318  \\ 
\hline \hline
trained on subset of NN50 & 0.196 & 0.175 \\ 
\hline
trained on subset of NN200 & 0.197 & 0.166 \\ 
\hline \hline
\end{tabular} 
\end{center}
\caption{Performance of ML models when confronted with data generated from
the elephant random walk (ERW). The two cases NN50, NN200 correspond to ERW,
when one takes every 50th or 200th data point. The table shows the accuracy
for the DL-based method introduced in \cite{SEC22} as well as a feature-based
method utilizing the features introduced in \cite{KOW22}. For reference the
table also shows what performance can be achieved, when the feature-based
model is trained on a subset of the ERW test data in the last two rows.}
\label{tab_elephant}
\end{table}

\begin{figure}
\centering
\includegraphics[width=0.9\linewidth]{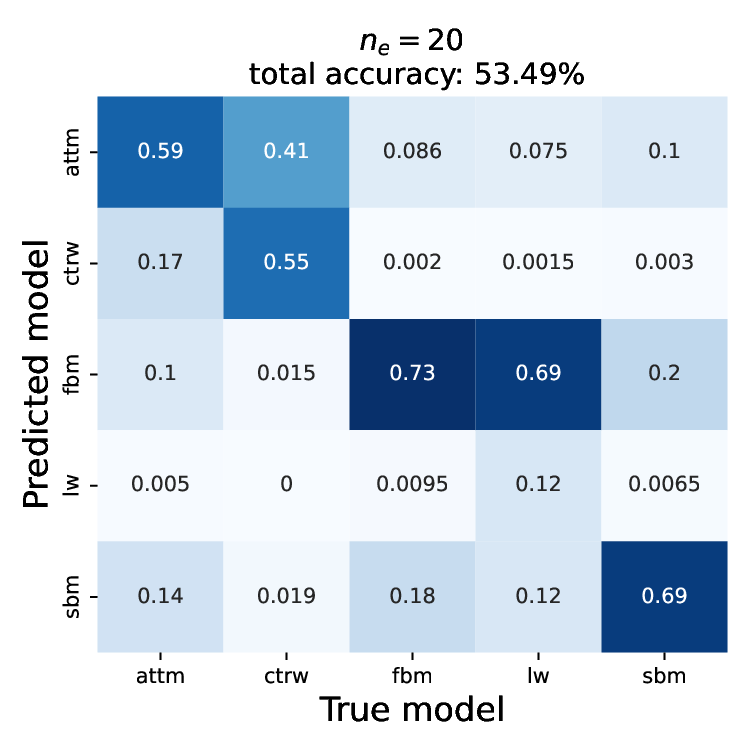}
\caption{Confusion matrix for dynamic noise with $n_e=20$ for the DL model
introduced in \cite{SEC22}. While FBM, ATTM and SBM show similar behavior
to the case without dynamic noise, identification of CTRW and especially LW
is strongly compromised by dynamic noise.}
\label{fig_dynconfmat}
\end{figure}

\begin{figure}
\centering
\includegraphics[width=0.7\linewidth]{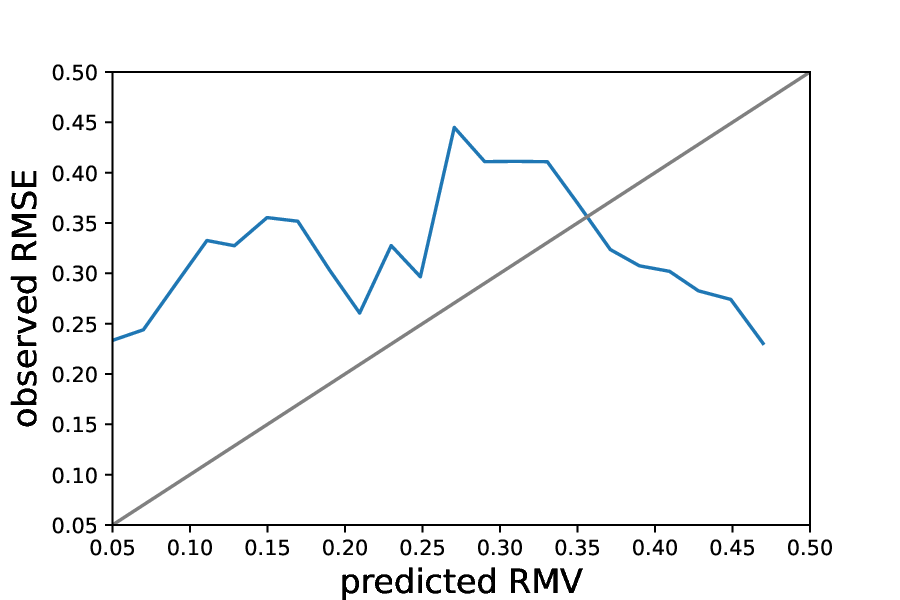}
\caption{Confidence Accuracy diagram obtained when trying to apply the DL
model from \cite{SEC22} to the elephant random walk. We see a strong deviation
between predicted root mean variance and observed root mean squared error,
indicating that the learning error prediction from the other models does
not translate well to the elephant random walk.}
\label{fig_elconf}
\end{figure}

During the AnDi-Challenge the competitors were provided with large training
data sets and tested on data generated from the same distributions as used for
the training. This practice gives an undeniable advantage to ML in general and
DL especially. Additionally the artificial data used in the AnDi-Challenge
only considered white Gaussian noise, which may not be sufficient to account
for all the noise sources present in experimental data.  To address these
problems we here test ML models to analyze their performance when confronted
with (a) data corrupted with dynamic noise and (b) the task of determining the
anomalous exponent for models not included in the training data. These tests
should indicate (a) how robust the methods are to different noise types and
(b) how well the learned determination of $\alpha$ can be generalized to
other models.

Dynamic noise stems from the finite exposure time needed to generate each
data point. In contrast to the additive Gaussian white noise this error
is characterized by temporal integration, which for discrete time steps is
replaced by a sum,
\begin{eqnarray}
\bar{x}(t)=\frac{1}{\tau_e}\int_0^{\tau_e} x({t-\xi})d\xi\to\frac{1}{n_e}
\sum_{j=0}^{n_e-1}x(t-j\Delta t),
\end{eqnarray}
where $\tau_e=n_e\Delta t$ is the exposure time consisting of $n_e$ time steps
of length $\Delta t$ \cite{SPO22,MEY23}. To test the ML models we generate
data sets, in the same manor as in the AnDi-Challenge, but with added dynamic
noise of different exposure times containing 10,000 trajectories each.
In table \ref{tab_dynnoise} we see the results when confronting the DL
model introduced in \cite{SEC22} and a feature-based model utilizing the
features from \cite{KOW22} with data corrupted by dynamic noise. For DL, the
determination of the anomalous exponent seems to be robust to the influence
of dynamic noise, only resulting in a slight performance decrease from an
MAE of $0.207$ to $0.235$ with increasing dynamic noise as characterized by
exposure time steps $n_e$, considering that a slight performance loss with
higher noise is to be expected. The model does however seem to struggle
with classification for high dynamic noise, where the accuracy drops
from $78\%$ down to $53.5\%$ for the highest considered dynamic noise
$n_e=20$. A look at the confusion matrix in figure \ref{fig_dynconfmat}
reveals that this is caused by miss-classification of LW and CTRW, evidently
the added Gaussian noise in the training data is not sufficient to account
for the changes incurred by dynamic noise.  The feature-based model proves
more robust to the influence of dynamic noise, showing a constant MAE of
$\approx0.22-0.23$, slightly outperforming the DL model for high dynamic
noise. Even more striking are the results obtained for classification,
while starting with a worse accuracy than DL ($\approx71\%$ compared to
$\approx78\%$), the feature-based model turns out to be much less hampered
by high dynamic noise levels, only decreasing the accuracy to $\approx65\%$
(compared to $\approx53.5\%$ for DL) at the highest noise level.  Critically,
when dealing with experimental setups with high dynamic noise, for accurate
classification, dynamic noise should therefore be included in the training
data sets, especially when relying on a DL model.

The elephant random walk (ERW) is a process with infinite memory, according
to which the next position of the walker is given by
\begin{equation}
x_i=x_{i-1}+\sigma_i,
\end{equation}
with the random variable $\sigma_i=\pm1$ \cite{SCH04}. The choice of $\sigma_i$
is determined through the memory of the previous time steps, by first drawing a
random integer $0\le j<i$ and then choosing $\sigma_i=\sigma_j$ with probability
$p$ or $\sigma_i=-\sigma_j$ with probability $1-p$. The first step $\sigma_0$ is
given as $\sigma_0=1$ with probability $q$ or $\sigma_0=-1$ with probability $1-
q$, for this work we choose $q=1/2$. In \cite{SCH04} it was shown that, in the
limit of many steps, this leads to the long time behavior of the MSD,
\begin{equation}
\langle x^2(t)\rangle\simeq\left\{\begin{array}{ll}
\frac{t}{3-4p},& p<3/4\\
t\ln(t),& p=3/4\\
\frac{t^{4p-2}}{(4p-3)\Gamma(4p-2)},& p>3/4,
\end{array}\right.   
\label{ltmsd}
\end{equation}
which corresponds to normal diffusion for $p<3/4$ and superdiffusion
for $p>3/4$ with $\alpha=4p-2$. We generate a data set containing
10,000 trajectories of length $T=100$, uniformly distributed in
$\alpha\in{1,1.05,\ldots,2.0}$, where for $\alpha=1$ we choose a $p<3/4$
randomly. To eliminate peculiarities caused by the constant step size,
and to provide sufficiently many steps to observe long time behavior
(\ref{ltmsd}) of the MSD, we only take every $NN=50$th or $NN=200$th data
point, effectively generating trajectories of length $5,000$ or $200,000$
and shortening them to length $100$, plus corrupting them with white Gaussian
noise.  The results when confronting the models with this data set are listed
in table \ref{tab_elephant}. For the DL model introduced in \cite{SEC22} with
this data set, we achieved an MAE of $\approx 0.246$ (for $NN=50$). While this
is a significant improvement from an unknown prediction of $\alpha\in[0.05,2]$
(MAE of $\approx0.49$), which the model was trained on, it does not improve
much on the performance expected when only identifying the elephant random
walk as superdiffusive with $\alpha\in[1,2]$ (MAE of $\approx 0.25$). In
addition, when considering the uncertainty predictions, that are provided by
the method as well, we find that the predictions learned on different models
are of little to no use when transferred to the elephant walk. Depicted
in figure \ref{fig_elconf}, we see that the predicted and observed errors
differ significantly. A possible reason for this can be found by closer
inspection of the predicted anomalous diffusion exponent, which reveals that
an unusually high number of trajectories are predicted at, or close to, a
ballistic motion with $\alpha=2$ ($31\%$ of trajectories are predicted with
$\alpha\ge1.9$ as compared to $9,5\%$ with a true $\alpha\ge1.9$). This might
be caused by the elephant random walk---on a single trajectory basis in the
superdiffusive regime---featuring a drift that only gets eliminated in the
ensemble average. Detecting this drift, while not specifically being trained
to deal with it, might lead the model to a falsely confident prediction of
ballistic motion. For the feature-based model in table \ref{tab_elephant}
we achieved only an MAE of $\approx0.348$ or $\approx0.318$ for $NN=50$
or $NN=200$ respectively. For reference we here also included the results
one can obtain when training the feature-based model on subsets of the
ERW data sets. These indicate what performance could be expected with
appropriately trained models achieving an MAE of $\approx0.196$ (for $NN=50$)
and $\approx0.166$ (for $NN=200$), and thereby significantly improving on the
performance of the models trained on the data sets of the AnDi-Challenge.
In conclusion we see that while some information can be extracted, for
accurate predictions, the machine needs to be trained on the appropriate model.

It should be noted however, that classical methods such as Bayesian Inference,
would similarly struggle when applied to a wrong prior, likely resulting
in falsely confident predictions as well. Likewise, it is known that not
considering a specific type of noise can lead to wrong predictions when
using statistical observables \cite{SPO22}.  Nevertheless, it is important
to stress that ML does not circumvent the necessity to consider such cases
and should be applied appropriately.

\paragraph*{Outlook.}

\begin{figure*}
\centering
\includegraphics[width=0.8\linewidth]{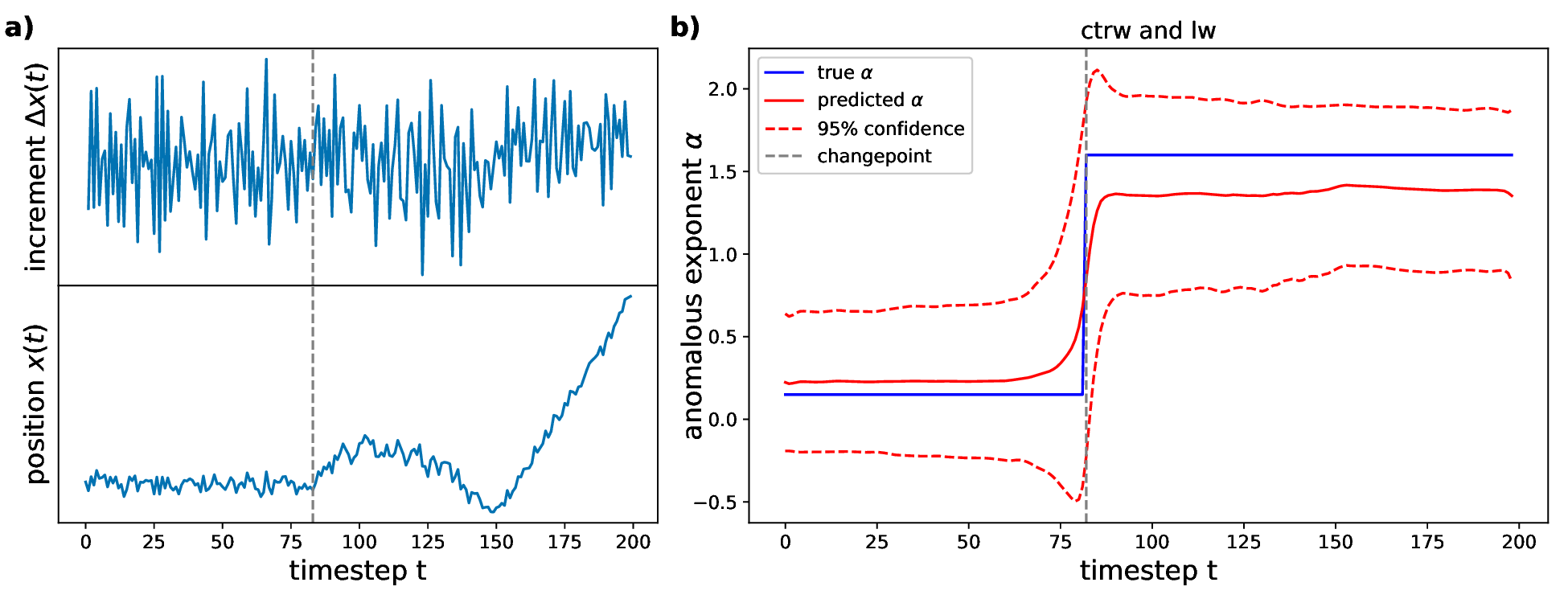}
\caption{Example of a sequence to sequence prediction for a trajectory with
a single change point. The trajectory shown in (a) changes from CTRW with
$\alpha=0.15$ to LW with $\alpha=1.60$ at time step $t_\text{change}=83$.
Panel (b) shows the output when feeding the case (a) into a trained neural
network. The used model includes an uncertainty estimation, whose 95\%
confidence interval is indicated by the red dashed line in the figure. Note
that the predicted error spikes at the change point.}
\label{fig_seqtoseq}
\end{figure*}

The results of the AnDi-Challenge proved the potential of ML approaches when
analyzing anomalous diffusion data. They come, however, at some price, often
acting as a Black Box, providing answers without explanation.  This lack of
explainability limits their usefulness when applied to real-world problems
and, inter alia, can lead to some overconfidence in the output results.
Building on the AnDi-Challenge we here presented two methods, which improve
the machine's explainability.

Extracting a set of statistical features, instead of using the raw position
data, allows us to use easier to interpret ML algorithms. In addition,
one can determine the importance of each feature, further improving
interpretability. In a recent publication, Mangalam et al.\ propose
multifractal features in order to improve the classification of anomalous
diffusion \cite{MAN23} which will be examined in future publications.
As an alternative to the trade-off brought by feature-based methods, we
can include an uncertainty prediction in the output of deep neural networks,
using the Multi-SWAG Bayesian DL method on top of the ML algorithm, at no cost
of accuracy. Apart from the obvious use of an added reliability estimate,
an analysis of these error predictions offers additional insights into the
learning process of the machine. Note that this method could also be applied to
feature-based DL approaches and that similar techniques for gradient boosting
or random forest algorithms exist \cite{MAL21}.  Despite these improvements,
we showed that these methods can still be hampered by out-of-distribution test
data, such as noise or models not included in the training data, possibly
leading to overconfident predictions due to the miss-specified prior. To
judge the validity of ML outputs and prepare appropriate training data sets,
analysis of experimental data using statistical methods remains necessary.
Moreover, we emphasize that visual inspection as well as some intuition on
the system will always present another layer of confidence, or caution.

Recent work has shown that sequence to sequence models are appropriate to
deal with trajectories changing between different diffusion models and/or
diffusion exponents \cite{REQ23,MAR23}, as was the target of the third task
of the AnDi-Challenge. Extending such models to include error estimates
will be a subject of future work. As an example we show the results of a
preliminary model, trained on trajectories with a single change point, in
figure \ref{fig_seqtoseq}. Apart from its use for uncertainty estimation,
the included error output can improve the extraction of change points from
the sequence, especially in cases where the anomalous diffusion exponents
before and after the change point are similar, inhibiting the determination
of change points only by means of the mean predictions.

Similarly to the example of superpositions of diffusion models used here,
in a recent work, Mu{\~n}oz-Gil et al.\ apply unsupervised learning to
anomalous diffusion, where different neural networks are trained to reproduce
trajectories generated from a specific diffusion model for each network. They
show that the differing performance in reproduction, when applied to different
diffusion models than trained on, can be used to classify a single, or a
superposition of, diffusion model(s) \cite{MUN21c}.

On another note, as we saw in figure \ref{fig_uncert_prob}, predictions on
very short trajectories tend to gravitate toward the center of the prior
distributions. This will limit the usefulness of single-trajectories
analysis when applied to experimental data consisting of many short
trajectories. Exploring the applicability of ML techniques to these kinds
of data may provide an interesting research avenue in the future, see also
the approach in \cite{CHE22}.

The application of ML, and its comparison to conventional methods, to
trajectory ensembles as well as trajectories with changing diffusion
models will be subject of the impending 2nd AnDi-Challenge.  In addition
this challenge will include video tracks of diffusing single particles,
without direct access to the positions of the tracers, thereby serving as
an exploration of noise-types different from the simple white Gaussian type,
inherent to the conversion from video tracks to particle trajectories.

Recent advances in computer vision could open a new track of research on
anomalous diffusion identification. The idea is quite simple: instead of
looking for custom neural network architectures for identification purposes
or preparing a robust set of features, one could (at least theoretically)
turn trajectories into images and feed them into well established pre-trained
computer vision models, that are known to excel in object recognition. The
main difficulty with this approach is that one cannot simply take a plot of
a trajectory as the image, since in this case the temporal structure of the
data is lost. Hence, one needs image representations of trajectories that
retain the existing spatial and temporal relations.

First approaches utilizing the computer-vision approach are very promising. For
instance, Garibo-i-Orts et al.~\cite{GAR23} used Gramian angular fields to
encode trajectories as images and two well-established pre-trained computer
vision models (ResNet and MobileNet) for both classification of diffusion types
and inference of the anomalous diffusion exponent $\alpha$. Their results for
short trajectories already outperform the state-of-the-art-methods. Markov
transition fields \cite{WAN15} or recurrence plots \cite{ECK87} are other
candidates for trajectory imaging methods that could potentially improve the
performance of the classifiers.  One of the benefits of the computer vision
approach is that it allows to use pre-trained models, which are available in
popular DL libraries like for instance \texttt{Keras}. In other words, it makes
the analysis accessible to researchers lacking an extensive background in ML.

\section*{Author information}

\paragraph*{Henrik Seckler} studied physics at the University of Potsdam,
Germany, where he is currently pursuing his doctorate. His research interests
include machine learning, data assimilation and anomalous diffusion.

\paragraph*{Janusz Szwabi{\'n}ski} obtained his BSc in Physics from
University of Wroc{\l}aw, Poland, and his PhD in Science from Saarland
University in Saarbruecken, Germany. He is currently a university professor
in the Department of Applied Mathematics at the Wroc{\l}aw University of
Science and Technology. He is mainly interested in complex systems and has a
track record in multidisciplinary research, with applications in statistical
physics, biology, social science and economy.

\paragraph*{Ralf Metzler} studied physics and obtained his PhD at
the University of Ulm, Germany. After postdoc positions at Tel Aviv
University and MIT he held faculty positions at NORDITA (Nordic Institute
for Theoretical Physics, Copenhagen, Denmark), University of Ottawa, Canada,
and Technical University of Munich, Germany. Ralf is chair professor in
Theoretical Physics at University of Potsdam. Among his distinctions, Ralf
received a Finland Distinguished Professorship from the Finnish Academy,
an Alexander von Humboldt Polish Honorary Research Scholarship, and the 2017
SigmaPhi Prize.  Ralf's research interest are in non-equilibrium statistical
physics and stochastic processes in complex systems, soft and bio matter,
and data assimilation.

\section*{Supporting Information}

Supporting Information: Definition of anomalous diffusion models, summary
of trajectory characteristics for the feature-based analysis, and
supplementary figures S1 and S2.

\section*{Acknowledgments}

HS thanks Timo J. Doerries for the implementation of the MIM model. JS
was supported by NCN-DFG Beethoven Grant No. 2016/23/G/ST1/04083. RM
acknowledges funding from the German Ministry for Education and Research
(NSF-BMBF project STAXS).

\section*{Competing interests}

The authors have no competing interests to declare.

\FloatBarrier

\clearpage

\title{Machine-Learning Solutions for the Analysis of Single-Particle
Diffusion Trajectories}

\author{Henrik Seckler}
\affiliation{Institute for Physics \& Astronomy, University of Potsdam, 14476
Potsdam-Golm, Germany}
\author{Janusz Szwabi{\'n}ski}
\affiliation{Hugo Steinhaus Center, Faculty of Pure and Applied Mathematics,
Wroc{\l}aw University of Science and Technology, Wybrze{\.z}e Wyspia{\'n}skiego
27, 50-370 Wroc{\l}aw, Poland}
\author{Ralf Metzler}\email{rmetzler@uni-potsdam.de}
\affiliation{Institute for Physics \& Astronomy, University of Potsdam, 14476
Potsdam-Golm, Germany}
\affiliation{{Asia Pacific Centre for Theoretical Physics, Pohang 37673,
Republic of Korea}}

\date{\today}

\clearpage

\onecolumngrid

\begin{center}
\textbf{\large Supplementary Material:\\[0.2cm]
Machine-Learning Solutions for the Analysis of Single-Particle
Diffusion Trajectories}\\[0.32cm]

Henrik Seckler,$^1$ Janusz Szwabi{\'n}ski,$^2$, and Ralf Metzler$^{1,3}$\\[0.2cm]
\textit{$^1$Institute for Physics \& Astronomy, University of Potsdam, 14476
Potsdam-Golm, Germany\\
$^2$Hugo Steinhaus Center, Faculty of Pure and Applied Mathematics,
Wroc{\l}aw University of Science and Technology, Wybrze{\.z}e Wyspia{\'n}skiego
27, 50-370 Wroc{\l}aw, Poland
$^3$Asia Pacific Centre for Theoretical Physics, Pohang 37673,
Republic of Korea}
\end{center}

\renewcommand{\thepage}{S\arabic{page}}
\renewcommand{\thefigure}{S\arabic{figure}}

\setcounter{page}{1}
\setcounter{figure}{0}

\appendix

\section{Anomalous Diffusion Models}
\label{sec_andi}

We here provide a brief primer on the considered anomalous diffusion models.

\paragraph*{CTRW.}
The continuous-time random walk (CTRW) is defined as a random walk, in
which the times between jumps and the spatial displacements are stochastic
variables \cite{MON65s,HUG81s,WEI89s}. In our case, we are considering a CTRW
for which the waiting time density $\psi(\tau)$ features a power law tail,
$\psi(\tau)\simeq\tau^{-1-\alpha}$ with scaling exponent $0<\alpha<1$. Thus
the characteristic waiting time diverges, $\int_0^\infty\tau\psi(\tau)d\tau
=\infty$. The spatial displacements follow a Gaussian law.

\paragraph*{LW.}
The L{\'e}vy walk (LW) is a special case of a CTRW. As above we consider
power-law distributed waiting times, $\psi(\tau)\simeq\tau^{-1-\sigma}$,
but the displacements are correlated, such that the walker always moves with
constant speed $v$ in one direction for one waiting time, randomly choosing
a new direction after each waiting time. One can show that this leads to an
anomalous diffusion exponent $\alpha$ given by \cite{ZAB15s}
\begin{equation}
\alpha=\left\{\begin{array}{ll}2&\mbox{ if } 0<\sigma<1 \mbox{ (ballistic
diffusion)}\\3-\sigma & \mbox{ if } 1<\sigma<2 \mbox{ (superdiffusion)}.
\end{array}\right. 
\end{equation}

\paragraph*{FBM.}
Fractional Brownian motion (FBM) is characterized by long-range, power-law
correlations between the increments. It is created by fractional Gaussian
noise for the increments, given by
\begin{equation}
\langle\xi_{fGn}(t)\xi_{fGn}(t+\tau)\rangle\sim\alpha(\alpha-1)K_\alpha\tau^{
\alpha-2}
\end{equation}
for sufficiently large $\tau$, where $\alpha$ is the anomalous diffusion
exponent and $K_\alpha$ is the generalised diffusion constant \cite{MAN68as}.

\paragraph*{SBM.}
Scaled Brownian motion (SBM) features the time dependent diffusivity $K(t)=
\alpha K_\alpha t^{\alpha-1}$, equivalent to the Langevin equation
\begin{equation}
\frac{dx(t)}{dt}=\sqrt{2K(t)}\xi(t),
\end{equation}
where $\xi(t)$ is white, zero-mean Gaussian noise \cite{JEO14s}.

\paragraph*{ATTM.}
Similar to SBM, the annealed transient time motion (ATTM) features a diffusion
coefficient $D$ varying over time. But in contrast to SBM, the change in
diffusivity is random in magnitude and occurs instantaneously in a manner
similar to the jumps in a CTRW. Here we consider diffusion coefficients
sampled from the density $P(D)\simeq D^{\sigma-1}$ and use a
delta-distributed waiting times $P(\tau)\propto \delta(\tau-D^{-\gamma})$,
with $\sigma<\gamma<\sigma+1$. As shown in \cite{MAS14s}, this leads
to subdiffusion with $\alpha=\sigma/\gamma$.

\section{Trajectory characteristics}
\label{sec_features}

In this appendix, the definitions of the features listed in table
1 are briefly introduced.

\subsection{Original features}

\subsubsection{Anomalous exponent}

Four estimates for the anomalous diffusion exponent $\alpha$ were used as
separate features:
\begin{enumerate}
\item the standard estimation, based on fitting the empirical TAMSD to
equation (2), 
\item 3 estimation methods proposed for trajectories with noise, which is
normally-distributed with zero mean \cite{LAN18s},
\begin{enumerate}
\item using the estimator
\begin{equation}
\hat{\alpha}=\frac{n_{\mathrm{max}} \sum_{n=1}^{n_{\mathrm{max}}}\ln (n)\ln(
\langle\mathbf{r}^2(n\Delta t)\rangle)-\sum_{n=1}^{n_{\mathrm{max}}}\ln(n)\left(
\sum_{n=1}^{n_{\mathrm{max}}}\ln(\langle\mathbf{r}^2(n\Delta t)\rangle) \right)}
{n_{\mathrm{max}}\sum_{n=1}^{n_{\mathrm{max}}}\ln^2(n)-\left(\sum_{n=1}^{
n_{\mathrm{max}}}\ln(n)\right)^2},
\end{equation}
with $n_{\mathrm{max}}$ equal to 0.1 times the trajectory length, rounded to
the nearest lower integer (but not less than 4),
\item simultaneous fitting of the parameters $\hat{D}$, $\hat{\alpha}$, and
$\hat{\sigma}$ in the relation
\begin{equation}
\langle\mathbf{r}^2(t)\rangle=2d\hat{D}t^{\hat{\alpha}}+\hat{\sigma}^2,
\end{equation}
where $d$ denotes the embedding dimension, $D$ is the diffusion coefficient,
and $\sigma^2$ is the variance of noise,
\item simultaneous fitting of the parameters $\hat{D}$ and $\hat{\alpha}$ in
the equation
\begin{equation}
\langle\mathbf{r}^2(n\Delta t)\rangle=2d\hat{D}(\Delta t)^{\hat{\alpha}}
(n^{\hat{\alpha}}-1).
\end{equation}
\end{enumerate}
\end{enumerate}

\subsubsection{Diffusion coefficient}

An estimator of the diffusion coefficient was extracted from the fit of the
empirical TAMSD to equation (2).

\subsubsection{Asymmetry}

The asymmetry is defined as 
\begin{equation}
A=-\log\left(1-\frac{(\lambda_1-\lambda_2)^2}{2(\lambda_1+\lambda_2)}\right),
\label{eq_asymmetry}
\end{equation}
where $\lambda_1$ and $\lambda_2$ are the eigenvalues of the tensor of gyration
\cite{SAX93s}. For a 2D random walk of $N$ steps, the tensor is given by
\begin{equation}
\mathbf{T} =\left(\begin{array}{cc}
\frac{1}{N+1}\sum_{j=0}^N(x_j -\langle x\rangle)^2 & \frac{1}{N+1}\sum_{j=0}^N
(x_j-\langle x\rangle)(y_j-\langle y\rangle)\\ 
\frac{1}{N+1}\sum_{j=0}^N(x_j-\langle x\rangle)(y_j-\langle y\rangle) &
\frac{1}{N+1}\sum_{j=0}^N(y_j-\langle y\rangle)^2
\end{array} 
\right),
\label{eq_tensor}
\end{equation} 
where $\langle x\rangle=(1/[N+1])\sum_{j=0}^N x_j$ is the average of the $x$
coordinate over all steps in the random walk. The asymmetry is expected to
help to detect directed motion.

\subsubsection{Efficiency}

The efficiency $E$ relates the net squared displacement of a particle to the
sum of squared step lengths,
\begin{equation}
E=\frac{|X_N-X_0|^2}{N\sum_{i=1}^N|X_i-X_{i-1}|^2}.
\label{eq_efficiency}
\end{equation} 
Similarly to the asymmetry, it may help to detect directed motion.

\subsubsection{Empirical velocity autocorrelation function}

The empirical velocity autocorrelation function \cite{WEB10s} for lag $1$ and
point $n$ is defined as
\begin{equation}
\chi_n=\frac{1}{N-n}\sum^{N-1-n}_{i=0}(X_{i+1+n}-X_{i+n})(X_{i+1}-X_{i}).
\end{equation}
It can be used to distinguish different subdiffusion processes. In \cite{KOW22s},
$\chi_n$ for points $n=1$ and $n=2$ was used.

\subsubsection{Fractal dimension}

The fractal dimension is a measure of the space-filling capacity of a pattern (a
trajectory in our case). For a planar trajectory, it may be calculated as 
\begin{equation}
D_f=\frac{\ln N}{\ln (NdL^{-1})},
\end{equation}
where $L$ is the total length of the trajectory, $N$ is the number of steps, and
$d$ is the largest distance between any two positions \cite{KAT85s}. It usually
takes values around $1$ for directed motion, around 2 for normal diffusion. For
subdiffusive CTRW it is also around 2, while for FBM it is larger than 2.

\subsubsection{Maximal excursion}

The maximal excursion of the particle is given by the formula
\begin{equation}
\textrm{ME}=\frac{\max(|X_{i+1}-X_{i})|}{X_N-X_0}
\end{equation}
It detects relatively long jumps (in comparison to the overall displacement).

\subsubsection{Mean maximal excursion}

The mean maximal excursion can replace the MSD as the observable used to
determine the anomalous diffusion exponent \cite{TEJ10s}. It is defined as
the standardized value of the largest distance traveled by a particle,
\begin{equation}
T_n=\frac{\max(|X_i-X_0|)}{\sqrt{\hat{\sigma}^2_N(t_N-t_0)}}.
\end{equation}
The parameter $\hat{\sigma}_N$ is a consistent estimator of the standard
deviation,
\begin{equation}
\hat{\sigma}^2_N=\frac{1}{2N\Delta t}\sum^N_{j=1}||X_j-X_{j-1}||^2_2.
\end{equation}

\subsubsection{Mean Gaussianity}

The Gaussianity $g(n)$ checks the Gaussian statistics of increments of a
trajectory \cite{ERN14s} and is defined as
\begin{equation}
g(n)=\frac{2\langle r_n^4\rangle}{3\langle r_n^2\rangle^2}-1,
\end{equation}
where $\langle r_n^k\rangle$ denotes the $k$th moment of the trajectory at time
lag $n$. The Gaussianity for normal diffusion is equal to 0. The same result
should be obtained for FBM, since its increments follow a Gaussian distribution.
Other types of motion should show deviations from that value.

Instead of looking at Gaussianities at single time lags, in \cite{KOW22s} the
mean Gaussianity over all lags was used as one of the features,
\begin{equation}
\langle g\rangle=\frac{1}{N}\sum^{N}_{i=1}g(n).
\end{equation}

\subsubsection{Mean-squared displacement ratio}

The MSD ratio gives information about the shape of the corresponding MSD curve.
We will define it as
\begin{equation}
MSDR(n_1,n_2)=\frac{\langle r_{n_1}^2\rangle}{\langle r_{n_2}^2\rangle}-
\frac{n_1}{n_2},
\label{eq_msdr}
\end{equation}
where $n_1<n_2$. $MSDR=0$ is zero for normal diffusion ($\alpha=1$). We should
get $MSDR\leq0$ for sub- and $MSDR\geq0$ for superdiffusion. In \cite{KOW22s},
$n_2=n_1+\Delta t$ was taken and then the averaged ratio over all $n_1=1,2,
\ldots,N-1$ was calculated for every trajectory.

\subsubsection{Kurtosis}

The Kurtosis gives insight into the asymmetry and peakedness of the distribution
of points within a trajectory \cite{HEL07s}. It is defined as the the fourth moment,
\begin{equation}
K=\frac{1}{N}\sum_{i=1}^N\frac{(x_i^p-\bar{x}^p)^4}{\sigma^4_{x^p}},\label{eq_kurtosis}
\end{equation}
of the projection of the position vectors $X_i$ onto the dominant eigenvector
$\vec{r}$ of the gyration tensor (\ref{eq_tensor}),
\begin{equation}
x_i^p=X_i\cdot\vec{r}.
\end{equation}
Here, $\bar{x}^p$ is the mean projected position and $\sigma_{x^p}$ the standard
deviation of $x^p$.

\subsubsection{Statistics based on $p$-variation}

The empirical $p$-variation is given by the formula \cite{BUR10s}
\begin{equation}
V_m^{(p)}=\sum_{k=0}^{\frac{N}{m}-1}|X_{(k+1)m}-X_{km}|^p\propto km^{\gamma^p}.
\end{equation}
This statistic can be used to detect fractional L\'{e}vy stable motion
(including FBM). Six features based on $V_m^{(p)}$ were used for the
classification of trajectories:
\begin{enumerate}
\item the power $\gamma^p$ fitted to $p$-variation for lags 1 to 5, 
\item the statistic $P$ used in \cite{LOC20s}, based on the monotonicity
changes of $V_m^{(p)}$ as a function of $m$:
\begin{equation}
P=\left\{\begin{array}{rl} 0 & \textrm{if $V_m^{(p)}$ does not change the
monotonicity},\\ 
1 & \textrm{if $V_m^{(p)}$ is convex for the highest $p$ for which it is not
monotonous},\\
-1 & \textrm{if $V_m^{(p)}$ is concave for the highest $p$ for which it is not
monotonous}.
\end{array} 
\right.
\end{equation}
\end{enumerate}

\subsubsection{Straightness}

The straightness $S$ measures the average direction change between subsequent
steps. It relates the net displacement of a particle to the sum of all step
lengths,
\begin{equation}
\label{eq_straightness}
S=\frac{|X_N-X_0|}{\sum_{i=1}^N|X_i-X_{i-1}|}. 
\end{equation}

\subsubsection{Trappedness}

The trappedness is understood as the probability that a diffusing particle is
trapped in a bounded region with radius $r_0$ up to some observation time $t$.
Saxton \cite{SAX93s} estimated this probability with
\begin{equation}
P(D,t,r_0)=10^{0.2048-2.5117(Dt/r_0^2)}.
\label{eq_trappedness}
\end{equation}
In the above expression, $r_0$ is approximated by half of the maximum distance
between any two positions along a given trajectory and $D$ is estimated by
fitting the first two points of the MSD curve (i.e., it is the so called
short-time diffusion coefficient).

\subsection{Additional features}

\subsubsection{D'Agostino-Pearson test statistic}

The d'Agostino-Pearson $\kappa^2$ test statistic \cite{AGO73s} measures the
departure of a given sample from normality. It is defined as
\begin{equation}
\kappa^2=Z_1(g_1)+Z_2(K),
\end{equation}
where $K$ is the sample kurtosis given by equation (\ref{eq_kurtosis}) and
$g_1=m_3/m_2^{3/2}$ is the sample skewness with $m_j$ being the $j$th sample
central moment. The transformations $Z_1$ and $Z_2$ should bring the
distributions of the skewness and kurtosis as close to the standard normal
as possible. This feature helps to distinguish ATTM and SBM from other motions.

\subsubsection{Kolmogorov-Smirnov statistic against $\chi^2$ distribution}

The Kolmogorov-Smirnov statistic quantifies the distance between the empirical
distribution function of the sample $F_n(X)$ and the cumulative distribution
function $G_n(X)$ of a reference distribution,
\begin{equation}
D_n=\sup_X|F_n(X)-G_n(X)|.
\end{equation}
Here, $n$ is the number of observations (i.e., the length of a trajectory).
The value of this statistic for the empirical distribution of squared
increments of a trajectory against the sampled $\chi^2$ distribution has been
taken as the next feature. The rationale of such choice is that for a Gaussian
trajectory the theoretical distribution of squared increments is the mentioned
$\chi^2$ distribution.

\subsubsection{Noah, Moses, and Joseph exponents}

For processes with stationary increments, there are in principle two mechanisms
that violate the Gaussian central limit theorem and produce anomalous scaling
of MSD: long-time increment correlations (referred to as the Joseph effect)
or a flat-tailed increment distribution (the Noah effect) \cite{AGH21s}. FBM
is the prototypical process that exhibits the first effect and LW the latter
one. An anomalous scaling can also be induced by a non-stationary increment
distribution \cite{AGH21s}. In this case, we talk about the Moses effect. It
should help to handle ATTM and SBM trajectories.

All three effects may be quantified by exponents, which can be used as
features. Given a stochastic process $X_t$ and the corresponding increment
process $\delta_t(\tau)=X_{t+\tau}-X_t$, the Joseph, Moses and Noah exponents
are defined as follows:
\begin{enumerate}
\item The Joseph exponent $J$ is estimated from the ensemble average of the
rescaled range statistic,
\begin{equation}
E\left[\frac{\max_{1\leq s\leq t}[X_s-\frac{s}{t}X_t]-\min_{1\leq s\leq t}[
X_s-\frac{s}{t}X_t]}{\sigma_t}\right]\sim t^J,
\end{equation}
where $\sigma_t$ is the standard deviation of the process $X_t$.
\item The Moses exponent $M$ is determined from the scaling of the ensemble
probability distribution of the sum of the absolute value of the increments which
can be estimated by the scaling of the median of the probability distribution
of $Y_t=\sum^{t-1}_{s=0}|\delta_s|$,
\begin{equation}
E[Y_t]\sim t^{M+\frac{1}{2}}.
\end{equation}
\item The Noah exponent $L$ is extracted from the scaling of the ensemble
probability distribution of the sum of increment squares, which can be estimated
by the scaling of the median of the probability distribution of $Z_t=\sum^{t-1}_{
s=0}\delta_s^2$:
\begin{equation}
E[Z_t]\sim t^{2L+2M-1}.
\end{equation}
\end{enumerate}

\subsubsection{Detrending moving average}

The detrending moving average (DMA) statistic \cite{BAL21s} is given by
\begin{equation}
DMA(\tau)=\frac{1}{N-\tau+1}\sum_{i=\tau}^N\left(X_i-\overline{X}^{\tau}_i
\right)^2,
\end{equation}
for $\tau=1,2,\ldots$, where $\overline{X}^{\tau}_i$ is a moving average of
$\tau$ observations, i.e., $\overline{X}^{\tau}_i=\frac{1}{\tau+1}\sum_{j=0}
^{\tau} X_{i-j}$. According to \cite{BAL21s}, a DMA-based statistical test
can help in the detection of the scaled Brownian motion. In \cite{KOW22s},
$DMA(1)$ and $DMA(2)$ were used as features.

\subsubsection{Average moving window characteristics}

Let us define the following moving window characteristic
\begin{equation}
MW=\frac{1}{2(N-m-1)}\sum_{t=1}^{N-m-1}\left|\textrm{sgn}\left(\overline{X}_{t+1}^{
(m)}-\overline{X}_t^{(m)}\right)-\textrm{sgn}\left(\overline{X}_{t}^{(m)}
-\overline{X}_{t-1}^{(m)}\right)\right|,
\end{equation}
where $\overline{X}^{(m)}$ denotes a statistic of the process calculated
within the window of length $m$ and $\textrm{sgn}$ is the sign function. In
\cite{KOW22s}, eight attributes based on the above formula have been used:
\begin{enumerate}
\item the mean and the standard deviation for $\overline{X}$ with windows of
lengths $m=10$ and $m=20$,
\item the $MW$ with the same window lengths calculated separately for $x$ and
$y$ coordinates.
\end{enumerate}

\subsubsection{Maximum standard deviation}

The last two features used in \cite{KOW22s} are based on the standard deviation
$\sigma_m$ of the process calculated within a window of length $m$. They are
given by
\begin{equation}
MXM=\frac{\max\left(\sigma_m(t)\right)}{\min\left(\sigma_m(t)\right)}
\end{equation}
and
\begin{equation}
MXC=\frac{\max\left|\sigma_m(t+1)-\sigma_m(t)\right|}{\sigma},
\end{equation}
where $\sigma$ denotes the overall standard deviation of the sample. A window
of length $m=3$ was used and the features were calculated for both coordinates separately. They should improve the detection of ATTM type of movements.

\newpage

\FloatBarrier

\section{Supplementary Figures}

\begin{figure*}[h]
\centering
\includegraphics[width=0.7\linewidth]{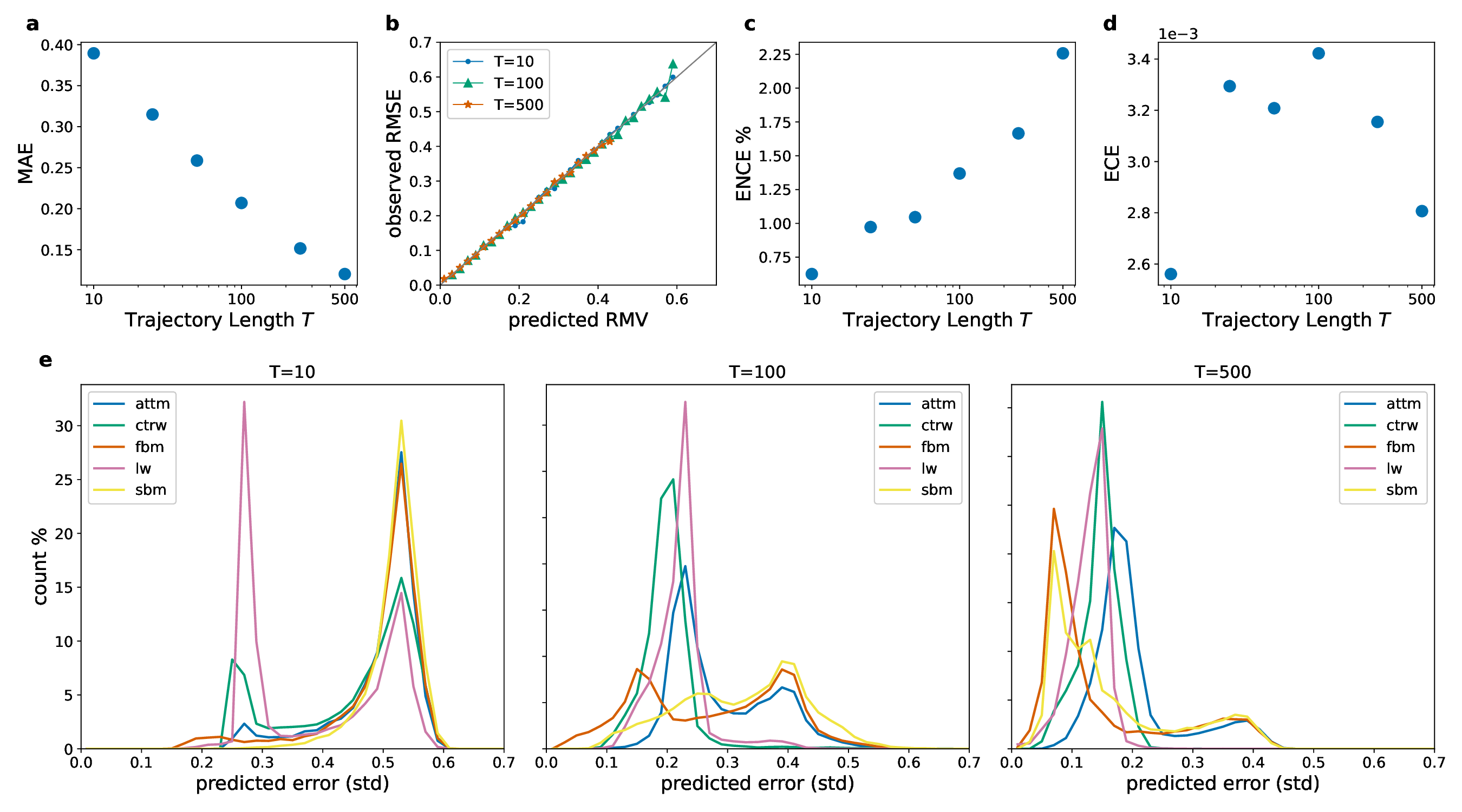}
\caption{Results, as presented in \cite{SEC22s}, for the regression of the 
anomalous diffusion exponent with included error estimation. The figure shows (a)
the regression performance via the mean absolute error in dependence of the 
trajectory length, (b) the reliability diagram, where coinciding predicted and 
observed error is represented as the diagonal, (c) the expected normalized (or 
unnormalized in (d)) calibration error, and (e) the predicted error distribution, 
for different trajectory length split depending on ground truth models. The 
distribution of errors in (e) has been extensively analyzed in \cite{SEC22s}, 
where it was used to gain insights into the learning process of the machine. For 
each trajectory length, the shown results were obtained from a test dataset of 
$10^5$ trajectories, using a model trained on $10^6$ trajectories of random 
models and exponents.}
\label{fig_regression}
\end{figure*}

\FloatBarrier
\newpage

\begin{figure*}
\centering
\includegraphics[width=0.7\linewidth]{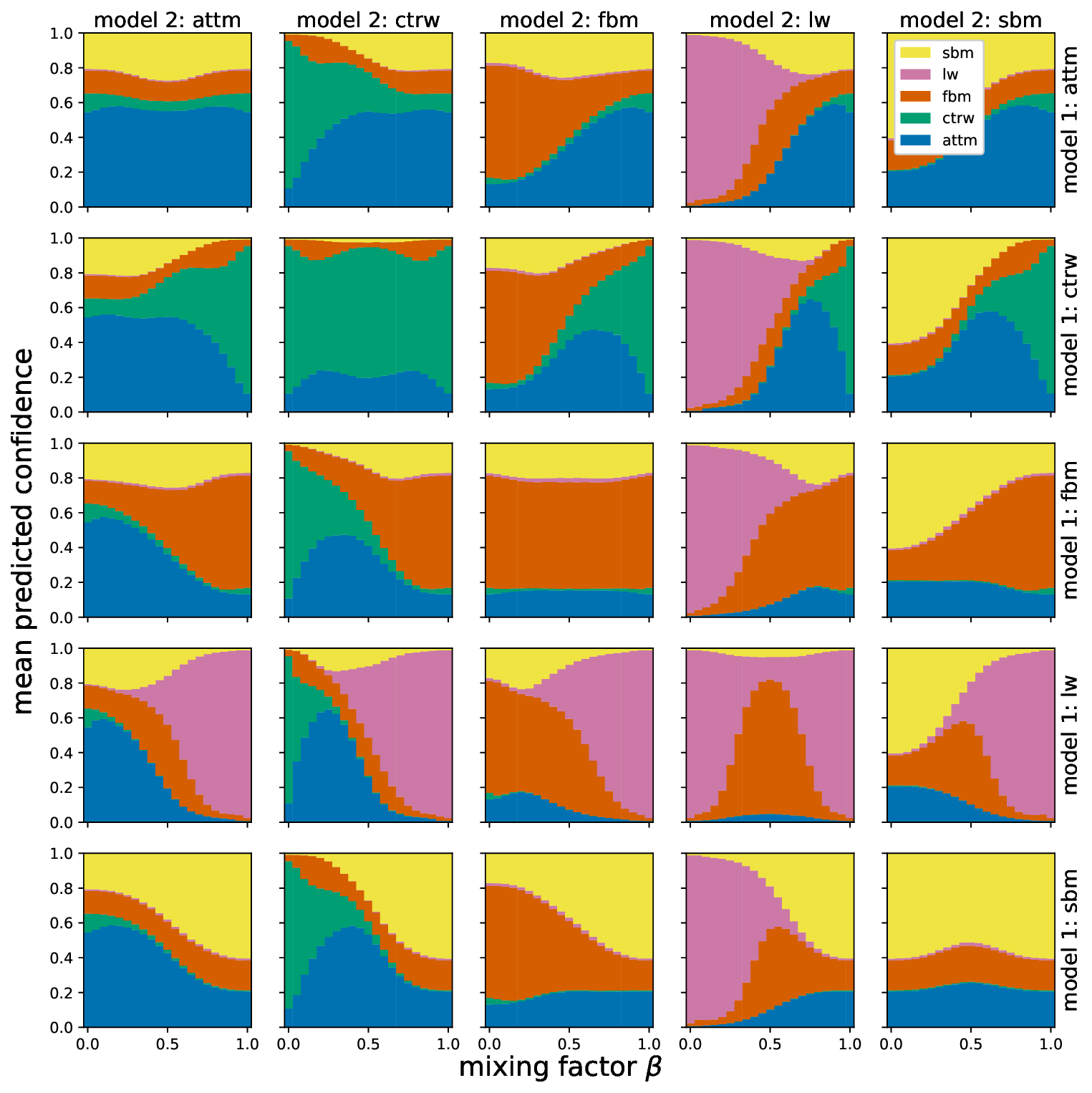}
\caption{Classification for a superposition of 2 models in 1D. The figure depicts 
the mean confidence assigned by the neural network, when given a mixture of two 
models in dependence of the superposition factor. The depicted results are 
obtained from 1D trajectories with 100 data points each.}
\label{fig_superpos_1d}
\end{figure*}


\begin{thebibliography}{999}
\section*{References}

\bibitem{BRO28} Brown, R.  brief account of microscopical observations 
made in the months of june, july and august 1827, on the particles contained 
in the pollen of plants; and on the general existence of active molecules in 
organic and inorganic bodies. \textit{Philos. Mag.} \textbf{1828}, 
\textit{4}, 161-173.

\bibitem{PER09} Perrin, J. Brownian movement and molecular reality. \textit{Ann. Chim. 
Phys.} \textbf{1909}, \textit{18}, 5-114.

\bibitem{ELF19} Elf, J.; Barkefors, I. Single-molecule
kinetics in living cells. \textit{Ann. Rev. Biochem.} \textbf{2019},
\textit{88}, 635-659.

\bibitem{CHE19} Cherstvy, A. G.; Thapa, S.; Wagner, C. E.; 
Metzler, R. Non-Gaussian, non-ergodic, and non-Fickian diffusion of
tracers in mucin hydrogels. \textit{Soft Matter} \textbf{2019}, 
\textit{15}, 2526-2551.

\bibitem{HOE13} H{\"o}fling, F.; Franosch, T. Anomalous
transport in the crowded world of biological cells, \textit{Rep. Prog. Phys.}
\textbf{2013}, \textit{76}, 046602.

\bibitem{HOR10} Horton, M. R.; H{\"o}fling, F.; R{\"a}dler, J. O.;
Franosch, T. Development of anomalous diffusion among crowding
proteins. \textit{Soft Matter} \textbf{2010}, \textit{6}, 2648-2656.

\bibitem{TOL04} Jeon, J.-H.; Tejedor, V.; Burov, S.; Barkai, E.; Selhuber-Unkel, C.; 
Berg-S{\o}rensen, K.; Oddershede, L.; Metzler, R. In vivo anomalous
diffusion and weak ergodicity breaking of lipid granules.
\textit{Phys. Rev. Lett.} \textbf{2011}, \textit{106}, 048103.

\bibitem{LEI12} Leijnse, N.; Jeon, J. H.; Loft, S.; Metzler, R.;
Oddershede, L. B. Diffusion inside living human cells. \textit{Eur. Phys. J.
Spec. Top.} \textbf{2012}, \textit{204}, 377a.

\bibitem{COD08} Codling, E. A.; Plank, M. J.; Benhamou, S. Random
walk models in biology. \textit{J. R. Soc. Interface} \textbf{2008}, \textit{5},
813-834.

\bibitem{spt} Gurtovenko, A. A.; Javanainen, M.; Lolicato, F.; Vattulainen, I.
The devil is in the details: what do we really track in single-particle tracking
experiments of diffusion in biological membranes?
\textit{J. Phys. Chem. Lett.} \textbf{2019}, \textit{10}, 1005-1011.

\bibitem{spt1} Roberts, T. D.; Yuan, R.; Xiang, L.; Delor, M.; Pokhrel, R.; Yang,
K.; Aqad, E.; Marangoni, T.; Trefonas, P.; Xu, K.; Ginsberg, N. S.
Direct correlation of single-particle motion to amorphous
microstructural components of semicrystalline poly(ethylene oxide)
electrolytic films.
\textit{J. Phys. Chem. Lett.} \textbf{2020}, \textit{11}, 4849-4858.

\bibitem{membr} Erimban, S.; Daschakraborty, S. Fickian yet non-Gaussian
nanoscopic lipid diffusion in the raft-mimetic membrane, \textit{J. Phys. Chem. B}
\textbf{2023}, \textit{127} 4939-4951.

\bibitem{membr1} Javanainen, M.; Martinez-Seara, H.; Metzler, R.; Vattulainen, I.
Diffusion of integral membrane proteins in protein-rich membranes.
\textit{J. Phys. Chem. Lett.} \textbf{2017}, \textit{8}, 4308-4313.

\bibitem{membr2} Winkler, P. M.; Regmi, R.; Garcia-Parajo, M.
Optical antenna-based fluorescence correlation spectroscopy to probe the
nanoscale dynamics of biological membranes. \textit{J. Phys. Chem. Lett.}
\textbf{2017}, \textit{9}, 110-119.

\bibitem{membr3} Spillane, K. M.; Ortega-Arroyo, J.; de Wit, G.; Eggeling, C.;
Ewers, H.; Wallace, M. I.; Kukura, P.
High-speed single-particle tracking of GM1 in model membranes reveals anomalous
diffusion due to interleaflet coupling and molecular pinning.
\textit{Nano Lett.} \textbf{2014}, \textit{14}, 5390-5397.

\bibitem{OKU86} Okubo, A. Dynamical aspects of animal grouping:
swarms, schools, flocks, and herds. \textit{Adv. Biophys.} \textbf{1986}, \textit{22}, 1-94.

\bibitem{VIL22a} Vilk, O.; Aghion, E.; Avgar, T.; Beta, C.; Nagel, O.;
Sabri, A.; Sarfati, R.; Schwartz, D. K.; Weiss, M.; Krapf, D.; Nathan, R.; Metzler, R.; 
Assaf, M. Unravelling the origins of anomalous diffusion: from
molecules to migrating storks. \textit{Phys. Rev. Res.} \textbf{2022}, \textit{4}, 033055.

\bibitem{BAR05} Bartumeus, F.; da Luz, M. G. E.; Viswanathan, G. M.; 
Catalan, J. Animal search strategies: a quantitative random-walk
analysis. \textit{Ecology} \textbf{2005}, \textit{86}, 3078-3087

\bibitem{ENG11} Engbert R; Mergenthaler, K; Sinn, P.; Pikovsky, A. An 
integrated model of fixational eye movements and microsaccades. \textit{Proc. 
Natl. Acad. Sci. U.S.A.} \textbf{2011}, \textit{108}, E765-E770.

\bibitem{eng} Herrmann, C. J. J.; Metzler, R.; Engbert, R. A self-avoiding walk
with neural delays as a model of fixational eye movements. \textit{Sci. Rep.}
\textbf{2017}, \textit{7}, 12958.

\bibitem{MAL99} Malkiel, B. G. A random walk down wall
street: including a life-cycle guide to personal investing. W. Norton \& Co:
New York, 1999.

\bibitem{PLE00} Plerou, V.; Gopikrishnan, P.; Amaral, L. A. N.; 
Gabaix, X.; Stanley, H. E. Economic fluctuations and anomalous
diffusion. \textit{Phys. Rev. E} \textbf{2000}, \textit{62}, R3023.

\bibitem{FER13} Fern{\'a}ndez, R.; Fr{\"o}hlich, J.; Sokal, A. D. 
Random walks, critical phenomena, and triviality in quantum field
theory. Springer Science \& Business Media: Berlin, 2013.

\bibitem{AND76} Anderson, J. B. Quantum chemistry by random walk.
H $^2\text{P}$, H$^+{}_3$ D$_{3 h}$ $^1\text{A'}_1$, H$_2$ $^3\Sigma^+{}_u$,
H$_4$ $^1\Sigma^+{}_g$, Be $^1\text{S}$. \textit{J. Chem. Phys.} \textbf{1976}, \textit{65}, 4121-4127.

\bibitem{LUD11} L{\"u}dtke, O.; Roberts, B. W.; Trautwein, U.; Nagy, G. 
A random walk down university avenue: life paths, life events, and
personality trait change at the transition to university life. \textit{J. Pers.
Soc. Psychol.} \textbf{2011}, \textit{101}, 620.

\bibitem{BOU03} Bouchaud, J.-P.; Potters, M. Theory of
financial risk and derivative pricing: from statistical physics to risk
management. Cambridge University Press: Cambridge, 2003.

\bibitem{PEA05} Pearson, K. The problem of the random walk.
\textit{Nature} \textbf{1905}, \textit{72}, 294.

\bibitem{MIS19} Mises, R. V. Fundamentals{\"a}tze
der Wahrscheinlichkeitsrechnung. \textit{Math. Z.} \textbf{1919}, \textit{4}, 1-97.

\bibitem{MON65} Montroll E. W.; Weiss, G. H. Random walks
on lattices. II. \textit{J. Math. Phys.} \textbf{1965}, \textit{6}, 167-181.

\bibitem{KAM92} Van Kampen, N. G. Stochastic processes
in chemistry and physics. Elsevier, 1992.

\bibitem{LEV48} L{\'e}vy, P. Processus stochastiques et
mouvement brownien. Gauthier-Villars: Paris, 1948.

\bibitem{HUG95} Hughes, B. D. Random walks and random
environments. Oxford University Press: Oxford, 1995.

\bibitem{EIN05} Einstein, A. {\"U}ber die von der
molekularkinetischen Theorie der W{\"a}rme geforderte Bewegung von
in ruhenden Fl{\"u}ssigkeiten suspendierten Teilchen. \textit{Ann. Phys.} 
\textbf{1905}, \textit{322}, 549-560.

\bibitem{SMO06} Von Smoluchowski, M. Zur kinetischen
Theorie der Brownschen Molekularbewegung und der Suspensionen. \textit{Ann. Phys.} 
\textbf{1906}, \textit{326}, 756-780.

\bibitem{SUT05} Sutherland, W. A dynamical theory of
diffusion for non-electrolytes and the molecular mass of albumin. \textit{Philos.
Mag.} \textbf{1905}, \textit{9}, 781-785.

\bibitem{LAN08} Langevin, P. Sur la th{\'e}orie du mouvement
brownien. \textit{C. R. Acad. Sci. (Paris)} \textbf{1908}, \textit{146}, 530-533.

\bibitem{BOU90} Bouchaud, J.-P.; Georges, A. Anomalous
diffusion in disordered media: statistical mechanisms, models and physical
applications. \textit{Phys. Rep.} \textbf{1990}, \textit{195}, 127-293.

\bibitem{MET00} Metzler, R.; Klafter, J. The random walk's
guide to anomalous diffusion: a fractional dynamics approach. \textit{Phys. Rep.} 
\textbf{2000}, \textit{339}, 1-77.

\bibitem{GOL06} Golding, I.; Cox, E. C. Physical
nature of bacterial cytoplasm. \textit{Phys. Rev. Lett.} \textbf{2006}, 
\textit{96}, 098102.

\bibitem{MAN15} Manzo, C.; Torreno-Pina, J. A.; Massignan, P.;
Lapeyre Jr., G. J.; Lewenstein, M.; Parajo, M. F. G. Weak ergodicity
breaking of receptor motion in living cells stemming from random diffusivity.
\textit{Phys. Rev. X} \textbf{2015}, \textit{5}, 011021.

\bibitem{KRA19} Krapf, D., Lukat, N.; Marinari, E.; Metzler, R.; Oshanin, G.;  
Selhuber-Unkel, C.; Selhuber-Unkel, A.; Stadler, L.; Weiss, M.; Xu; X. Spectral 
content of a single non-Brownian trajectory, \textit{Phys. Rev. X} 
\textbf{2019}, \textit{9}, 011019.

\bibitem{STA17} Stadler, L.; Weiss, M. Non-equilibrium
forces drive the anomalous diffusion of telomeres in the nucleus of mammalian
cells. \textit{New J. Phys.} \textbf{2017}, \textit{19}, 113048.

\bibitem{KIN17} Kindermann, F.; Dechant, A.; Hohmann, M.;
Lausch, T.; Mayer, D.; Schmidt, F.; Lutz, E.; Widera, A. Nonergodic
diffusion of single atoms in a periodic potential. \textit{Nat. Phys.} \textbf{2017},
\textit{13}, 137-141.

\bibitem{SOK12} Sokolov, I. M. Models of anomalous diffusion in
crowded environments. \textit{Soft Matter} \textbf{2012}, \textit{8}, 9043-9052.

\bibitem{SAX94} Saxton, M. J. Anomalous diffusion due to obstacles:
a Monte Carlo study. \textit{Biophys. J.} \textbf{1994}, \textit{66}, 394-401.

\bibitem{SAX01} Saxton, M. J. Anomalous subdiffusion in fluorescence
photobleaching recovery: a monte carlo study. \textit{Biophys. J.} \textbf{2001}, \textit{81}, 2226-2240.

\bibitem{BUR11} Burov, S.; Jeon, J. H.; Metzler, R.; Barkai, E. Single
particle tracking in systems showing anomalous diffusion: the role of weak
ergodicity breaking. \textit{Phys. Chem. Chem. Phys.} \textbf{2011}, \textit{13}, 1800-1812.

\bibitem{ERN14} Ernst, D.; K{\"o}hler, J.; Weiss, M. Probing the
type of anomalous diffusion with single-particle tracking. \textit{Phys. Chem.
Chem. Phys.} \textbf{2014}, \textit{16}, 7686-7691.

\bibitem{anna} Bodrova, A.; Chechkin, A. V.; Cherstvy, A. G.; Metzler, R.
Quantifying non-ergodic dynamics of force-free granular gases. \textit{Phys.
Chem. Chem. Phys.} \textbf{2015}, \textit{17}, 21791-21798.

\bibitem{anna1} Bodrova, A.; Chechkin, A. V.; Cherstvy, A. G.; Safdari, H.;
Sokolov, I. M.; Metzler, R.
Underdamped scaled Brownian motion: (non-)existence of the overdamped
limit in anomalous diffusion. \textit{Sci. Rep.} \textbf{2016}, \textit{6},
30520.

\bibitem{LIM02} Lim, S. C.; Muniandy, S. V. Self-similar Gaussian
processes for modeling anomalous diffusion. \textit{Phys. Rev. E} \textbf{2002}, 
\textit{66}, 021114.

\bibitem{JEO14} Jeon, J.-H.; Chechkin, A. V.; Metzler, R. Scaled
Brownian motion: a paradoxical process with a time dependent diffusivity for
the description of anomalous diffusion. \textit{Phys. Chem. Chem. Phys.} \textbf{2014},
\textit{16}, 15811-15817.

\bibitem{MAN68a} Mandelbrot, B. B.; Van Ness, J. W.
Fractional Brownian motions, fractional noises and applications.
\textit{SIAM Rev.} \textbf{1968}, \textit{10}, 422-437.

\bibitem{MET14} Metzler, R.; Jeon, J. H.; Cherstvy, A. G.;
Barkai, E. Anomalous diffusion models and their properties:
non-stationarity, non-ergodicity, and ageing at the centenary of
single particle tracking. \textit{Phys. Chem. Chem. Phys.} \textbf{2014}, 
\textit{16}, 24128-24164.

\bibitem{MAR22} Marcone, B.; Nampoothiri, S.; Orlandini, E.; Seno, F.; 
Baldovin, F. Brownian non-Gaussian diffusion of self-
avoiding walks. \textit{J. Phys. A} \textbf{2022}, \textit{55}, 354003.

\bibitem{HEG22} Hegde, A. S.; Chandrashekar, C. M.
Characterization of anomalous diffusion in one-dimensional 
quantum walks. \textit{J. Phys. A} \textbf{2022}, \textit{55}, 234006.

\bibitem{VIT22} Vitali, S.; Paradisi, P.; Pagnini, G. 
Anomalous diffusion originated by two Markovian hopping-trap
mechanisms. \textit{J. Phys. A} \textbf{2022}, \textit{55}, 224012.

\bibitem{SAB22} Sabzikar, F.; Kabala, J.; Burnecki, K.; 
Tempered fractionally integrated process with stable noise as 
a transient anomalous diffusion model. \textit{J. Phys. A} \textbf{2022}, 
\textit{55}, 174002.

\bibitem{WAN20} Wang, W.; Cherstvy, A. G.; Chechkin, A. V.; Thapa, S.;
Seno, F.; Liu, X.; Metzler, R. Fractional Brownian motion 
with random diffusivity: emerging residual nonergodicity below the 
correlation time. \textit{J. Phys. A} \textbf{2020}, \textit{53}, 474001.

\bibitem{HUG81} Hughes, B. D.; Shlesinger, M. F.; Montroll, E. W.
Random walks with self-similar clusters. \textit{Proc. Natl. Acad. Sci.
U.S.A.} \textbf{1981}, \textit{78}, 3287-3291.

\bibitem{WEI89} Weissman, H.; Weiss, G. H.; Havlin, S.
Transport properties of the continuous-time random walk with a
long-tailed waiting-time density. \textit{J. Stat. Phys.} \textbf{1989}, 
\textit{57}, 301-317.

\bibitem{LEV37} L{\'e}vy, P. Th{\'e}orie de l'addition des
variables al{\'e}atoires. Gauthier-Villars: Paris, 1937.

\bibitem{CHE08} Chechkin, A. V.; Metzler, R.; Klafter, J.; 
Gonchar, V. Y. Introduction to the theory of L{\'e}vy flights. 
In \textit{Anomalous Transport: Foundations and Applications}; 
Klages, R.; Radons, G.; Sokolov, I. M., Eds.;
Springer: Berlin, 2008; pp 129-162.

\bibitem{SHL86} Shlesinger, M. F.; Klafter, J.
L{\'e}vy walks versus L{\'e}vy flights, on growth and form.
Springer: Dordrecht, 1986.

\bibitem{ZAB15} Zaburdaev, V.; Denisov, S.; Klafter, J.
L{\'e}vy walks. \textit{Rev. Mod. Phys.} \textbf{2015}, \textit{87}, 
483.

\bibitem{MAG20} Magdziarz, M.; Zorawik, T. Limit 
properties of L{\'e}vy walks. \textit{J. Phys. A} \textbf{2020}, \textit{53}, .

\bibitem{MAS14} Massignan, P.; Manzo, C.; Torreno-Pina, J. A.;
Garc{\'i}a-Parajo, M. F.; Lewenstein, M.; Lapeyre Jr., G. J. Nonergodic
subdiffusion from Brownian motion in an inhomogeneous medium. \textit{Phys. Rev.
Lett.} \textbf{2014}, \textit{112}, 150603.

\bibitem{MER15} Meroz, Y.; Sokolov, I. M. A toolbox for
determining subdiffusive mechanisms. \textit{Phys. Rep.} \textbf{2015}, 
\textit{573}, 1-29.

\bibitem{MAK11} Makarava, N.; Benmehdi, S.; Holschneider, M.
Bayesian estimation of self-similarity exponent. \textit{Phys. Rev. E} 
\textbf{2011}, \textit{84}, 021109.

\bibitem{MET09} Metzler, R.; Tejedor, V.; Jeon, J. H.; He, Y.; 
Deng, W. H.; Burov, S.; Barkai, E. Analysis of single particle
trajectories: from normal to anomalous diffusion. \textit{Acta Phys. Pol. B}
\textbf{2009}, \textit{40}.

\bibitem{MAG09} Magdziarz, M.; Weron, A.; Burnecki, K.; Klafter, J. 
Fractional Brownian motion versus the continuous-time random walk: a
simple test for subdiffusive dynamics. \textit{Phys. Rev. Lett.}
\textbf{2009}, \textit{103}, 180602.

\bibitem{MET19} Metzler, R. Brownian motion and beyond:
first-passage, power spectrum, non-Gaussianity, and anomalous diffusion. 
\textit{J. Stat. Mech.} \textbf{2019}, \textit{2019}, 114003.

\bibitem{VIL22b} Vilk, O.; Aghion, E.; Nathan, R.; Toledo, S.; Metzler, R.; 
Assaf, M. Classification of anomalous diffusion in animal movement
data using power spectral analysis. \textit{J. Phys. A} \textbf{2022}, 
\textit{55}, 334004.

\bibitem{SPO22} Sposini, V.; Krapf, D.; Marinari, E.; et al., Towards a robust 
criterion of anomalous diffusion. \textit{Commun. Phys.} \textbf{2022},
\textit{5}, 305.

\bibitem{CON07} Condamin, S.; B{\'e}nichou, O.; Tejedor, V.;
Voituriez, R; Klafter, J. First-passage times in complex
scale-invariant media. \textit{Nature} \textbf{2007},
\textit{450}, 77-80.

\bibitem{SLE19} Slezak, J.; Metzler, R.; Magdziarz, M.
Codifference can detect ergodicity breaking and non-Gaussianity.
\textit{New J. Phys.} \textbf{2019}, \textit{21}, 053008.

\bibitem{MAR20} Maraj, K.; Szarek, D.; Sikora, G.; Wy{\l}oma{\'n}ska, A. 
empirical anomaly measure for finite-variance processes. 
\textit{J. Phys. A} \textbf{2020}, \textit{54}, 024001.

\bibitem{CHE17} Chen, L.; Bassler, K. E.; McCauley, J. L.; 
Gunaratne, G. H. Anomalous scaling of stochastic 
processes and the Moses effect. \textit{Phys. Rev. E}
\textbf{2017}, \textit{95}, 042141.

\bibitem{MAR02} Martin, D. S.; Forstner, M. B.; K{\"a}s, J. A. 
Apparent subdiffusion inherent to single particle tracking. 
\textit{Biophys. J.} \textbf{2002}, \textit{83}, 2109-2117.

\bibitem{PRE20} Prezhdo, O. V. Advancing physical chemistry with machine 
learning. \textit{J. Phys. Chem. Lett.} \textbf{2020}, \textit{11}, 9656-9658.

\bibitem{BAC19} Back, S.; Yoon, J.; Tian, N.; Zhong, W.; Tran, K.; Ulissi, Z. W. 
Convolutional neural network of atomic surface structures to predict binding 
energies for high-throughput screening of catalysts. \textit{J. Phys. Chem. 
Lett.} \textbf{2019}, \textit{10}, 4401-4408.

\bibitem{BAT20} Batra, R.; Chan, H.; Kamath, G.; Ramprasad, R.; Cherukara, M. J.; 
Sankaranarayanan, S. Screening of therapeutic agents for COVID-19 using 
machine learning and ensemble docking studies. \textit{J. Phys. Chem. Lett.}, 
\textbf{2020}, \textit{11}, 7058-7065.

\bibitem{DRA20} Dral, P. O. Quantum chemistry in the age of machine learning. 
\textit{J. Phys. Chem. Lett.} \textbf{2020}, \textit{11}, 2336-2347.

\bibitem{GRA19} Granik, N.; Weiss, L. E.; Nehme, E.; Levin, M.; Chein, M.;
Perlson, E.; Roichman, Y.; Shechtman, Y. Single-particle diffusion
characterization by deep learning. \textit{Biophys. J.} \textbf{2019},
\textit{117}, 185-192.

\bibitem{BO19} Bo, S.; Schmidt, F.; Eichhorn, R.; Volpe, G. Measurement
of anomalous diffusion using recurrent neural networks. \textit{Phys. Rev. E}
\textbf{2019}, \textit{100}, 010102.

\bibitem{MUN20b} Mu{\~n}oz-Gil, G.; Garcia-March, M. A.;
Manzo, C.; Mart{\'i}n-Guerrero, J. D.; Lewenstein, M. Single
trajectory characterization via machine learning. \textit{New J. Phys.} 
\textbf{2020}, \textit{22}, 013010.

\bibitem{MUN20a} Mu{\~n}oz-Gil, G.; Volpe, G.;
Garc{\'i}a-March, M. A.; Metzler, R.; Lewenstein, M.; Manzo, C. The
anomalous diffusion challenge: single trajectory characterisation as a
competition. In \textit{Emerging Topics in Artificial Intelligence 2020}, 
Vol. 11469. SPIE, 2020.

\bibitem{andijpa} Manzo, C.; Mu{\~n}oz-Gil, G.; Volpe, G.; Garcia-March, M. A.;
Lewenstein, M.; Metzler, R. Preface: Characterisation of physical processes
from anomalous diffusion data. \textit{J. Phys. A} \textbf{2023}, \textit{56},
010401.

\bibitem{GAJ21} Gajowczyk, M.; Szwabi\'nski, J. 
Detection of anomalous diffusion with deep residual networks. 
\textit{Entropy} \textbf{2021}, \textit{23}, 649.

\bibitem{BUR10} Burnecki, K.; Weron, A. {Fractional 
L{\'e}vy stable motion can model subdiffusive dynamics}.
\textit{Phys. Rev. E} \textbf{2010}, \textit{82}, 021130.

\bibitem{LOC20} Loch-Olszewska, H.; Szwabi{\'n}ski, J.
Impact of feature choice on machine learning classification of
fractional anomalous diffusion. \textit{Entropy} \textbf{2020}, 
\textit{22}, 1436.

\bibitem{noise} Jeon, J.-H.; Barkai, E.; Metzler, R. Noisy continuous time
random walks. \textit{J. Chem. Phys.} \textbf{2013}, \textit{139}, 121916.

\bibitem{MAN68b} Mandelbrot B. B.; Wallis, J. R. Noah, Joseph,
and operational hydrology. Water Resour. Res. \textbf{1968}, 
\textit{4}, 909-918.

\bibitem{AGH21} Aghion, E.; Meyer, P. G.; Adlakha, V.; Kantz, H.; Bassler, K. E.
Moses, Noah and Joseph effects in L{\'e}vy walks. \textit{New J. Phys.}
\textbf{2021}, \textit{23}, 023002.

\bibitem{MEY22} Meyer, P. G.; Aghion, E.; Kantz, H. Decomposing the
effect of anomalous diffusion enables direct calculation of the Hurst
exponent and model classification for single random paths. 
\textit{J. Phys. A} \textbf{2022}, \textit{55}, 274001.

\bibitem{THA18} Thapa, S.; Lomholt, M. A.; Krog, J.; Cherstvy, A. G.; 
Metzler, R. Bayesian analysis of single-particle tracking data using 
he nested-sampling algorithm: maximum-likelihood model selection applied 
to stochastic-diffusivity data. \textit{Phys. Chem. Chem. Phys.} \textbf{2018},
\textit{20}, 29018-29037.

\bibitem{PAR21} Park, S.; Thapa, S.; Kim, Y.; Lomholt, M. A.; Jeon, J.-H.
Bayesian inference of L{\'e}vy walks via hidden Markov models. \textit{J.
Phys. A} \textbf{2021}, \textit{54}, 484001.

\bibitem{SZE14} Szegedy, C.; Zaremba, W.; Sutskever, I.;
Bruna, J.; Erhan, D.; Goodfellow, I.; Fergus, R. Intriguing
properties of neural networks. In \textit{2nd International 
Conference on Learning Representations}, ICLR 2014.

\bibitem{KOW19} Kowalek, P.; Loch-Olszewska, H.; Szwabi\'nski, J. 
Classification of diffusion modes in single-particle tracking data: 
feature-based versus deep-learning approach. \textit{Phys. Rev. E}
\textbf{2019}, \textit{100}, 032410.

\bibitem{MUN21a} Mu{\~n}oz-Gil, G.; Volpe, G.;
Garc{\'i}a-March, M. A.; et al.\, Objective comparison of methods to
decode anomalous diffusion. \textit{Nat. Commun.} \textbf{2021}, \textit{12}, 6253.

\bibitem{ARG21} Argun, A.; Volpe, G.; Bo, S. Classification,
inference and segmentation of anomalous diffusion with recurrent neural
networks. \textit{J. Phys. A} \textbf{2021}, \textit{54}, 294003.

\bibitem{GAR21} Garibo-i-Orts, {\`O}.; Baeza-Bosca, A.; Garcia-March, A.;
Conejero, J. A. Efficient recurrent neural network methods for
anomalously diffusing single particle short and noisy trajectories. 
\textit{J. Phys. A} \textbf{2021}, \textit{54}, 504002.

\bibitem{LI21} Li, D.; Yao, Q.; Huang, Z. WaveNet-based deep neural
networks for the characterization of anomalous diffusion (WADNet). 
\textit{J. Phys. A} \textbf{2021}, \textit{54}, 404003.

\bibitem{FIR23} Firbas, N.; Garibo-i-Orts, {\`O}.; Garcia-March, M. {\'A}.; 
Conejero, J. A. Characterization of anomalous diffusion through
convolutional transformers. \textit{J. Phys. A} \textbf{2023},
\textit{56}, 014001.

\bibitem{ALH22} AL-hada, E. A.; Tang, X.; Deng, W.
Classification of stochastic processes by convolutional neural
networks. \textit{J. Phys. A} \textbf{2022}, \textit{55}, 274006.

\bibitem{GEN21} Gentili A.; Volpe, G. Characterization of
anomalous diffusion classical statistics powered by deep learning (CONDOR).
\textit{J. Phys. A} \textbf{2021}, \textit{54}, 314003,

\bibitem{MAN21} Manzo, C. Extreme learning machine for the
characterization of anomalous diffusion from single trajectories (andi-elm).
\textit{J. Phys. A} \textbf{2021}, \textit{54}, 334002.

\bibitem{KOW22} Kowalek, P.; Loch-Olszewska, H.; {\L}aszczuk, {\L}.;
Opa{\l}a, J.; Szwabi{\'n}ski, J. Boosting the performance of anomalous
diffusion classifiers with the proper choice of features. \textit{J. Phys. A}
\textbf{2022}, \textit{55}, 244005.

\bibitem{PIN21} Pinholt, H. D.; Bohr, S. S. R.; Iversen, J. F.;
Boomsma, W.; Hatzakis, N. S. Single-particle diffusional fingerprinting:
a machine-learning framework for quantitative analysis of heterogeneous
diffusion. \textit{Proc. Natl. Acad. Sci. U.S.A.}
\textbf{2021}, \textit{118}, e2104624118.

\bibitem{SEC22} Seckler, H.; Metzler, R. Bayesian deep learning 
for error estimation in the analysis of anomalous diffusion. 
\textit{Nat. Commun.} \textbf{2022}, \textit{13} 6717.

\bibitem{THA22} Thapa, S.; Park, S.; Kim, Y.; Jeon, J. H.; Metzler, R.;
Lomholt, M. A. Bayesian inference of scaled versus fractional Brownian
motion. \textit{J. Phys. A} \textbf{2022}, \textit{55}, 194003.

\bibitem{MUN21b} Mu{\~n}oz-Gil, G.; Requena, B.; Volpe, G.; Garcia-March, M.A.;
Manzo, C. \textit{AnDiChallenge/ANDI\_datasets: Challenge 2020 release (v.1.0)},  
\url{https://doi.org/10.5281/zenodo.4775311}, (accessed 2023-07-24).

\bibitem{KRO18} Krog, J.; Jacobsen, L. H.; Lund, F. W.; W{\"u}stner, D.;
Lomholt, M. A. Bayesian model selection with fractional Brownian
motion. \textit{J. Stat. Mech.} \textbf{2018}, \textit{2018}, 093501.

\bibitem{JAN20} Janczura, J.; Kowalek, P.; Loch-Olszewska, H.;
Szwabi{\~n}ski, J.; Weron, A. Classification of particle trajectories
in living cells: machine learning versus statistical testing hypothesis for
fractional anomalous diffusion. \textit{Phys. Rev. E}
\textbf{2020}, \textit{102}, 032402.

\bibitem{VER21} Verdier, H.; Duval, M.; Laurent, F.; Cass{\'e}, A.;
Vestergaard, C. L.; Masson, J. B. Learning physical properties of
anomalous random walks using graph neural networks. 
\textit{J. Phys. A} \textbf{2021}, \textit{54}, 234001.

\bibitem{VER22} Verdier, H.; Laurent, F.; Cass{\'e}, A.; Vestergaard, C. L.; 
Masson, J. B. Variational inference of fractional Brownian 
motion with linear computational complexity. \textit{Phys. Rev. E}
\textbf{2022}, \textit{106}, 055311.

\bibitem{LEC15} LeCun, Y.; Bengio, Y.; Hinton, G. Deep learning,.
\textit{Nature} \textbf{2015}, \textit{521}, 436-444.

\bibitem{CHE22} Chen, Z; Geffroy, L.; Biteen, J. nobias: Analyzing 
anomalous diffusion in single-molecule tracks with nonparametric Bayesian
inference. \textit{Front. Bioinform.} \textbf{2021}, \textit{1}, 742073.

\bibitem{FUK80} Fukushima, K. neocognitron: A self-organizing neural 
network model for a mechanism of pattern recognition unaffected by shift 
in position. \textit{Biol. Cybern.} \textbf{1980}, \textit{36}, 193-202.

\bibitem{HOC97} Hochreiter S.; Schmidhuber, J. Long
short-term memory. \textit{Neural Comput.} \textbf{1997}, \textit{9}, 1735-1780.

\bibitem{KIU09} Kiureghian A.; Ditlevsen, O. Aleatory
or epistemic? Does it matter?. \textit{Structural Safety} \textbf{2009},
\textit{31}, 105-112.

\bibitem{KEN17} Kendall, A.; Gal, Y. What
uncertainties do we need in Bayesian deep learning for computer vision?.
\textit{Adv. Neural Inf. Process. Syst.} \textbf{2017}, \textit{30}, 5580-5590.

\bibitem{WAN22} Wang, Q.; Ma, Y.; Zhao, K.; Tian, Y. A
comprehensive survey of loss functions in machine learning. 
\textit{Ann. Data Sci.} \textbf{2022}, \textit{9}, 1-26.

\bibitem{NIX94} Nix, D. A.; Weigend, A. S. Estimating the mean
and variance of the target probability distribution. In \textit{Proc. 1994 IEEE Int.
Conf. Neural Networks (ICNN'94)}, Vol. 1. IEEE, 1994.

\bibitem{KOL50} Kolmogorov, A. N.; Foundations of the
theory of probability. Chelsea Publishing Co.: London, 1950.

\bibitem{MET49} Metropolis, N.; Ulam, S. The Monte Carlo
method. \textit{J. Am. Stat. Assoc.} \textbf{1949},
\textit{44}, 335-34.

\bibitem{binder} Binder, K.; Monte Carlo methods in statistical physics.
Springer: Berlin, 1986.

\bibitem{LAK17} Lakshminarayanan, B.; Pritzel, A.;
Blundell, C. Simple and scalable predictive uncertainty estimation
using deep ensembles. \textit{Adv. Neural Inf. Process. Syst.} \textbf{2017},
\textit{30}.

\bibitem{GAL16a} Gal, Y.; Ghahramani, Z. Dropout as a Bayesian
approximation: representing model uncertainty in deep learning. In \textit{Int. Conf.
Machine Learning} PMLR, 2016.

\bibitem{GAL16b} Gal, Y. Uncertainty in deep learning.
Doctoral dissertation. (Cambridge University: Cambridge, 2016).

\bibitem{MAD19} Maddox, W. J.; Izmailov, P.; Garipov, T.; Vetrov, D. P.;
Wilson, A. G. A simple baseline for Bayesian uncertainty in
deep learning. \textit{Adv. Neural Inf. Process. Syst.} \textbf{2019}, 
\textit{32}.

\bibitem{WIL20} Wilson, A. G.; Izmailov, P. Bayesian
deep learning and a probabilistic perspective of generalization. 
\textit{Adv. Neural Inf. Process. Syst.} \textbf{2020}, \textit{33}, 4697-4708.

\bibitem{BOT10} Bottou, L. Large-scale machine learning with
stochastic gradient descent. In \textit{Proceedings of COMPSTAT'2010: 
19th International Conference on Computational StatisticsParis 
France, August 22-27, 2010 Keynote, Invited and Contributed Papers}. 
Physica-Verlag HD, 2010, 177-186.

\bibitem{NAE15} Naeini, M. P.; Cooper, G.; Hauskrecht, M.
Obtaining well calibrated probabilities using Bayesian binning.
In \textit{Proceedings of the AAAI conference on artificial intelligence.} 
Vol. 29. No. 1. 2015.

\bibitem{LEV20} Levi, D.; Gispan, L.; Giladi, N.; Fetaya, E.
Evaluating and calibrating uncertainty prediction in regression
tasks. \textit{Sensors} \textbf{2022}, \textit{22}, 5540.

\bibitem{DOE22} Doerries, T. J.; Chechkin, A. V.; Metzler, R. 
Apparent anomalous diffusion and non-Gaussian distributions 
in a simple mobile-immobile transport model with Poissonian switching. 
\textit{J. R. Soc. Interface} \textbf{2022}, \textbf{19}, 20220233.

\bibitem{GEN76} Van Genuchten, M. T.; Wierenga, P. J. Mass
transfer studies in sorbing porous media
I. analytical solutions. \textit{Soil Sci. Soc. Am. J.} \textbf{1976}, 
\textit{40}, 473-480.

\bibitem{DEA63} Deans, H. H. A mathematical model for
dispersion in the direction of flow of porous
media. \textit{Soc. Pet. Eng. J.} \textbf{1963}, \textit{3}, 49-52.

\bibitem{WAG17} Wagner, T.; Kroll, A.; Haramagatti, C. R.; Lipinski, H.-G.; 
Wiemann, M. Classification and segmentation of nanoparticle 
diffusion trajectories in cellular micro environments. \textit{PLoS ONE} 
\textbf{2017}, \textit{12}, e0170165.

\bibitem{WER19} Weron A.; Janczura, J.; Boryczka, E.; Sungkaworn, T.; 
Calebiro, D. {Statistical testing approach for fractional 
anomalous diffusion classification.} \textit{Phys. Rev. E} \textbf{2019},
\textit{99}, 042149.

\bibitem{JAM13} James, G.; Witten, D.; Hastie, T.; Tibshirani, R. 
{An introduction to statistical learning with applications in R.} 
Springer: New York, 2013.

\bibitem{SON15} Song, Y.-Y.; LU, Y. {Decision tree methods: 
applications for classification and prediction.}, \textit{Shanghai Archives of 
Psychiatry} \textbf{2015}, \textit{27}, 130.

\bibitem{MEY23} Meyer, P. G.; Metzler, R. Stochastic processes in a confining 
harmonic potential in the presence of static and dynamic measurement noise. 
\textit{New J. Phys.} \textbf{2023}, \textit{25}, 063003

\bibitem{SCH04} Sch{\"u}tz, G. M.; Trimper, S. Elephants can always remember: 
Exact long-range memory effects in a non-Markovian random walk. \textit{Phys. 
Rev. E} \textbf{2004}, \textit{70}, 045101.

\bibitem{MAN23} Mangalam, M.; Metzler, R.; 
Kelty-Stephen, D. G. {Ergodic characterization of 
non-ergodic anomalous diffusion processes.} 
\textit{Phys. Rev. Res.} \textbf{2023}, \textit{5}, 023144.

\bibitem{MAL21} Malinin, A.; Prokhorenkova, L.; 
Ustimenko, A. Uncertainty in gradient boosting via ensembles.
International Conference on Learning Representations \textbf{2020};
E-print arXiv:2006.10562.

\bibitem{REQ23} Requena, B.; Mas{\'o}, S.; Bertran, J.;
Lewenstein, M.; Manzo, C.; Mu{\~n}oz-Gil, G. {Inferring 
pointwise diffusion properties of single trajectories with deep 
learning}. arXiv:2302.00410 (2023).

\bibitem{MAR23} Martinez, Q.; Chen, C.; Xia, J.; Bahai, H. 
{Sequence-to-sequence change-point detection in 
single-particle trajectories via recurrent neural network 
for measuring self-diffusion}. \textit{Transp. Porous Media} \textbf{2023},
\textit{147}, 679-701.

\bibitem{MUN21c} Mu{\~n}oz-Gil, G.; i Corominas, G. G.; 
Lewenstein, M. {Unsupervised learning of anomalous diffusion
data: an anomaly detection approach}. \textit{J. Phys. A} 
\textbf{2021}, \textit{54}, 504001.

\bibitem{GAR23} Garibo-i-Orts, \`O.; Firbas, N.; 
Sebasti\'a, L.; Conejero, J. A. {Gramian angular 
fields for leveraging pretrained computer vision models with 
anomalous diffusion trajectories}. \textit{Phys. Rev. E}
\textbf{2023}, \textit{3}, 034138.

\bibitem{WAN15} Wang, Z.; Oates, T. 
{Imaging time-series to improve classification and imputation}.
In \textit{Proceedings of the 24th International Conference on 
Artificial Intelligence.} 2015.

\bibitem{ECK87} Eckmann, J.-P.; Oliffson Kamphorst, S.; Ruelle,  D.
{Recurrence plots of dynamical systems}. 
\textit{EPL} \textbf{1987}, \textit{4}, 973.

\end{thebibliography}

\begin{thebibliography}{100}


\bibitem{MON65s} Montroll E. W.; Weiss, G. H. Random walks
on lattices. II. \textit{J. Math. Phys.} \textbf{1965}, \textit{6}, 167-181.

\bibitem{HUG81s} Hughes, B. D.; Shlesinger, M. F.; Montroll, E. W.
Random walks with self-similar clusters. \textit{Proc. Natl. Acad. Sci.
U.S.A.} \textbf{1981}, \textit{78}, 3287-3291.

\bibitem{WEI89s} Weissman, H.; Weiss, G. H.; Havlin, S.
Transport properties of the continuous-time random walk with a
long-tailed waiting-time density. \textit{J. Stat. Phys.} \textbf{1989}, 
\textit{57}, 301-317.

\bibitem{ZAB15s} Zaburdaev, V.; Denisov, S.; Klafter, J.
L{\'e}vy walks. \textit{Rev. Mod. Phys.} \textbf{2015}, \textit{87}, 
483.

\bibitem{MAN68as} Mandelbrot, B. B.; Van Ness, J. W.
Fractional Brownian motions, fractional noises and applications.
\textit{SIAM Rev.} \textbf{1968}, \textit{10}, 422-437.

\bibitem{JEO14s} Jeon, J.-H.; Chechkin, A. V.; Metzler, R. Scaled
Brownian motion: a paradoxical process with a time dependent diffusivity for
the description of anomalous diffusion. \textit{Phys. Chem. Chem. Phys.} \textbf{2014},
\textit{16}, 15811-15817.

\bibitem{MAS14s} Massignan, P.; Manzo, C.; Torreno-Pina, J. A.;
Garc{\'i}a-Parajo, M. F.; Lewenstein, M.; Lapeyre Jr., G. J. Nonergodic
subdiffusion from Brownian motion in an inhomogeneous medium. \textit{Phys. Rev.
Lett.} \textbf{2014}, \textit{112}, 150603.

\bibitem{LAN18s} Lanoisel\'ee, Y.; Sikora, G.; Grzesiek, A.; Grebenkov, D. S.; 
Wy\l{}oma\'nska,  A. {Optimal parameters for 
anomalous-diffusion-exponent estimation from noisy data}. 
\textit{Phys. Rev. E} \textbf{2018}, \textit{98}, 062139.

\bibitem{SAX93s} Saxton, M. J. {Lateral diffusion in an 
archipelago. single-particle diffusion}. \textit{Biophys. J.} \textbf{1993}, 
\textit{64}, 1766-1780.

\bibitem{WEB10s} Weber, S. C.; Spakowitz, A. J.; Theriot, J. A. 
{Bacterial chromosomal loci move subdiffusively through 
a viscoelastic cytoplasm}. \textit{Phys. Rev. Lett.} \textbf{2010},
\textit{104}, 238102.

\bibitem{KOW22s} Kowalek, P.; Loch-Olszewska, H.; {\L}aszczuk, {\L}.;
Opa{\l}a, J.; Szwabi{\'n}ski, J. Boosting the performance of anomalous
diffusion classifiers with the proper choice of features. \textit{J. Phys. A}
\textbf{2022}, \textit{55}, 244005.

\bibitem{KAT85s} Katz, M. J.; George, E. B. {Fractals 
and the analysis of growth paths}. \textit{Bull. Math. Biol.}
\textbf{1985}, \textit{47}, 273-286.

\bibitem{TEJ10s} Tejedor, V.; B{\'e}nichou, O.; Voituriez, R.; Jungmann, R.; 
Simmel, F.; Selhuber-Unkel, C.; Oddershede, L. B.; Metzler, R. 
{Quantitative analysis of single particle trajectories: 
mean maximal excursion method}. \textit{Biophys. J.}
\textbf{2010}, \textit{98}, 1364-1372.

\bibitem{ERN14s} Ernst, D.; K{\"o}hler, J.; Weiss, M. Probing the
type of anomalous diffusion with single-particle tracking. \textit{Phys. Chem.
Chem. Phys.} \textbf{2014}, \textit{16}, 7686-7691.

\bibitem{HEL07s} Helmuth, J. A.; Burckhardt, C. J.; Koumoutsakos, P.; 
Greber, U. F.; Sbalzarini, I. F. {A novel supervised 
trajectory segmentation algorithm identifies distinct types of 
human adenovirus motion in host cells}. \textit{J. Struct. Biol.} 
\textbf{2007}, \textit{159}, 347-358.

\bibitem{BUR10s} Burnecki, K.; Weron, A. {Fractional 
L{\'e}vy stable motion can model subdiffusive dynamics}.
\textit{Phys. Rev. E} \textbf{2010}, \textit{82}, 021130.

\bibitem{LOC20s} Loch-Olszewska, H.; Szwabi{\'n}ski, J.
Impact of feature choice on machine learning classification of
fractional anomalous diffusion. \textit{Entropy} \textbf{2020}, 
\textit{22}, 1436.

\bibitem{AGO73s} D'Agostino, R.; Pearson, E. S. {Tests for 
departure from normality. Empirical results for the distributions 
of b2 and $\sqrt{b1}$}. \textit{Biometrika} \textbf{1973}, 
\textit{60}, 613-622.

\bibitem{AGH21s} Aghion, E.; Meyer, P. G.; Adlakha, V.; Kantz, H.; Bassler, K. E.
Moses, Noah and Joseph effects in L{\'e}vy walks. \textit{New J. Phys.}
\textbf{2021}, \textit{23}, 023002.

\bibitem{BAL21s} Balcerek, M.; Burnecki, K.; Sikora, G.; 
Wy\l{}oma\'nska, A. {Discriminating Gaussian processes 
via quadratic form statistics}. \textit{Chaos} \textbf{2021},
\textit{31}, 063101.

\bibitem{SEC22s} Seckler, H.; Metzler, R. Bayesian deep learning 
for error estimation in the analysis of anomalous diffusion. 
\textit{Nat. Commun.} \textbf{2022}, \textit{13} 6717.

\end{thebibliography}
\end{document}